\documentclass[journal]{IEEEtran}

\usepackage{capjelcommands}

\providecommand{\herm}{\mathrm{H}} 
\providecommand{\trans}{\mathrm{T}}

\providecommand{\JBSL}[2]{\mathrm{j}_{#1} \left( #2 \right)}  
\providecommand{\YBSL}[2]{\mathrm{y}_{#1} \left( #2 \right)}
\providecommand{\Zvac}{Z_0}

\providecommand{\Prad}{P_\mathrm{r}}
\providecommand{\WSTO}{\widetilde{W}_\mathrm{sto}}
\providecommand{\WE}{\widetilde{W}_\mathrm{e}}
\providecommand{\WM}{\widetilde{W}_\mathrm{m}}
\providecommand{\We}{W_\mathrm{e}}
\providecommand{\Wm}{W_\mathrm{m}}
\providecommand{\XD}{\M{X}'}
\providecommand{\XE}{\widetilde{\M{X}}_\mathrm{e}}
\providecommand{\XM}{\widetilde{\M{X}}_\mathrm{m}}
\providecommand{\Ivec}{\M{I}} 
\providecommand{\Vi}{\M{V}}
\providecommand{\iopt}{\mathrm{opt}}

\providecommand{\Nmeas}{N_\mathrm{n}}  
\providecommand{\OmegaN}[1]{\widetilde{\Omega}^{#1}}  

\providecommand{\Qmin}{Q_\mathrm{min}}
\providecommand{\QT}{Q} 
\providecommand{\QU}{Q_\mathrm{U}}
\providecommand{\Qext}{Q_\mathrm{ext}}
\providecommand{\QchuTM}{Q_\mathrm{Chu}^{\mathrm{{TM}}}}
\providecommand{\QchuTMTE}{Q_\mathrm{Chu}^{\mathrm{{TMTE}}}}

\newcommand\figwidth{8.8} 
\newcommand\tabwidth{1.8} 

\begin{document}
\title{Optimal Composition of Modal Currents For Minimal Quality Factor Q}
\author{Miloslav~Capek,~\IEEEmembership{Member,~IEEE,}
        and~Lukas~Jelinek
\thanks{Manuscript received March 20, 2016; revised XXX. This work was supported by the Czech Science Foundation under project No.~15-10280Y and by the Technology Agency of the Czech Republic, grant TA 04010457.}
\thanks{The authors are with the Department of Electromagnetic Field, Faculty of Electrical Engineering, Czech Technical University in Prague, Technicka 2, 16627, Prague, Czech Republic
(e-mail: miloslav.capek@fel.cvut.cz).}
}

\markboth{}%
{Capek \MakeLowercase{\textit{et al.}}: Optimal Composition of Modal Currents For Minimal Quality Factor Q}
\maketitle

\begin{abstract}
This work describes a powerful, yet simple, procedure how to acquire a current approaching the lower bound of quality factor $Q$. This optimal current can be determined for an arbitrarily shaped electrically small radiator made of a perfect conductor. Quality factor~$Q$ is evaluated by Vandenbosch's relations yielding stored electromagnetic energy as a function of the source current density. All calculations are based on a matrix representation of the integro-differential operators. This approach simplifies the entire development and results in a straightforward numerical evaluation. The optimal current is represented in a basis of modal currents suitable for solving the optimization problem so that the minimum is approached by either one mode tuned to the resonance, or, by two properly combined modes. An overview of which modes should be selected and how they should be combined is provided and results concerning rectangular plate, spherical shell and fractal shapes of varying geometrical complexity are presented. The reduction of quality factor~$Q$ and the $G$/$Q$ ratio are studied and, thanks to the modal decomposition, the physical interpretation of the results is discussed in conjunction with the limitations of the proposed procedure.
\end{abstract}

\begin{IEEEkeywords}
Antenna theory, electromagnetic theory, modal analysis, Q factor.
\end{IEEEkeywords}

%
\IEEEpeerreviewmaketitle

\section{Introduction}
\label{Intro}
\IEEEPARstart{Q}{uality} factor $Q$ is a parameter of primary importance for electrically small radiators \cite{VolakisChenFujimoto_SmallAntennas} due to inverse proportionality to the fractional bandwidth, FBW, \cite{Sievenpiper_ExpretimentalValidationOfPerformanceLimitsAndDesignGuidelines}. In comparison to the direct evaluation or measurement of the FWB, the calculation of quality factor $Q$ has many appealing properties as described in \cite{VolakisChenFujimoto_SmallAntennas}. 

The most favourable feature of the quality factor~$Q$ probably lies in its ability to determine the fundamental bounds delimiting the maximal available FBW. These limits are considered without any particular feeding and they are predetermined only by the given shape of a radiator. Knowing the bounds allows us to understand how good, in principle, a radiator could be. See \cite{Sievenpiper_ExpretimentalValidationOfPerformanceLimitsAndDesignGuidelines} for a comprehensive summary of all relevant work on quality factor~$Q$.

The history of seeking the fundamental bounds of quality factor~$Q$ is equally long and exciting, referring back to Chu \cite{Chu_PhysicalLimitationsOfOmniDirectAntennas} and Wheeler \cite{Wheeler_FundamentalLimitationsOfSmallAntennas}. Many authors have struggled with a canonical example of a spherical shell and, thanks to the propitious symmetries of spherical geometry, the problem has been successfully solved analytically, see, among others, \cite{Chu_PhysicalLimitationsOfOmniDirectAntennas, CollinRotchild_EvaluationOfAntennaQ, Fante_QFactorOfGeneralIdeaAntennas, Thal_ExactCircuitAnalysisOfSphericalWaves, McLean_AReExaminationOfTheFundamentalLimitsOnTheRadiationQofESA, HansenCollin_ANewChuFormulaForQ}. Nevertheless, the area allocated for the design of a radiator is, in practice, always strictly prescribed and it rarely occupies a spherical region. Consequently, the question of fundamental bounds of quality factor~$Q$ for an arbitrarily shaped radiator naturally arises.

Before we delve into the in-depth analysis, the optimal current and the optimal antenna must be distinguished carefully. The optimal current fills the prescribed region and yields extremal values of optimized parameter(s) \cite{Gustafsson_OptimalAntennaCurrentsForQsuperdirectivityAndRP, GustafssonTayliEhrenborgCismasuNorbedo_MatlabCvxTutorial}. How the current is excited or supported by a conducting platform is neither guaranteed nor specified. On the other hand, the optimal antenna is composed of a conducting body and its operation is governed by a feeding network \cite{Sievenpiper_ExpretimentalValidationOfPerformanceLimitsAndDesignGuidelines, BestHanna_AperformanceComparisonOfFundamentalESA}. While the optimal current poses a useful theoretical concept, the optimal antenna represents more realistic prototype. 

In terms of ESA design, the privileged problem when dealing with the optimal currents is the minimization of quality factor~$Q$ for an arbitrarily shaped region. Unfortunately, this task forms a non-convex problem \cite{NocedalWright_NumericalOptimization, BoydVandenberghe_ConvexOptimization} which, until now, has only been treated by heuristic methods \cite{RahmatMichielssen_ElectromagneticOptimizationByGenetirAlgorithms, RahmatSamii_Kovitz_Rajagopalan-NatureInspiredOptimizationTechniques} or by a relaxed technique \cite{Kim_LowerBoundsOnQForFinizeSizeAntennasOfArbitraryShape, GustafssonTayliEhrenborgCismasuNorbedo_MatlabCvxTutorial}. By focusing not directly on quality factor~$Q$ but merely on the $G/Q$ ratio, the problem of minimization the $G/Q$ can be solved in an efficient way via polarizabilities \cite{GustafssonSohlKristensson_PhysicalLimitationsOfAntennasOfArbitraryShape_RoyalSoc, GustafssonCismasuJonsson_PhysicalBoundsAndOptimalCurrentsOnAntennas_TAP}, by convex optimization \cite{Gustafsson_OptimalAntennaCurrentsForQsuperdirectivityAndRP, GustafssonTayliEhrenborgCismasuNorbedo_MatlabCvxTutorial}, or even on a circuit level \cite{Thal_QboundsForArbitrarySmallAntennas}. However, regardless of the sustained effort \cite{JonssonGustafsson_StoredEnergiesInElectricAndMagneticCurrentDensities_RoyA}, the problem of the minimal quality factor~$Q$ is still not completely resolved and clearly there is room for further investigation. 

The performance of the optimal current can be approached by optimal antennas and this has been approved for spherical \cite{Best_TheRadiationPropertiesOfESAsphericalHelix, Best_LowQelectricallySmallLinearAndEllipticalPolarizedSphericalDipoleAntennas, BestHanna_AperformanceComparisonOfFundamentalESA} and also for planar rectangular \cite{Best_ElectricallySmallResonantPlanarAntennas} regions. The most general approach for isolating optimal antennas involves a powerful heuristic optimization, operating almost exclusively over source currents, \cite{CismasuGustafsson_FBWbySimpleFreuqSimulation, CismasuGustafsson_MultibandAntennaQoptimizationUsingStoredEnergyExpressions, HassanWadbroBerggren_TopologyOptimizationOfMetallicAntennas, HassanNorelandAugustineWadbroBerggren_TopologyOptimizationOfPlanarAntennaForWidebandNearFieldCoupling, YangAdams_SystematicShapeOptimizationOfSymmetricMIMOAntennasUsingCM}. The favoured algorithm to extract optimal antennas is the so-called pixelling technique \cite{JorhsonRahmatSamii_GAandMoMforTheDesignOfIntegratedAntennas}. Being aware of the realistic limits justifies the considerable effort in recent research on optimal currents.

This paper describes an alternative procedure for approaching the fundamental bounds of quality factor~$Q$ and the $G$/$Q$ ratio of an arbitrarily shaped radiator whose electrical dimensions are small. Idea of how to tune the current into the resonance by an additional current is first generally elaborated without any particular solution of minimum quality factor~$Q$, then the optimization problem is expressed in terms of modal currents. While this idea is not completely new \cite{HarringtonMautz_ControlOfRadarScatteringByReactiveLoading, ChalasSertelVolakis_ComputationOfTheLimitsForASAusingCM, GustafssonTayliEhrenborgCismasuNorbedo_MatlabCvxTutorial}, it is finally resolved in this paper, including the closely related numerical aspects. 

Additional physical insight, primarily obtained due to the proper modal decomposition, makes it possible to reduce the complexity of the problem immensely. The optimization procedure is then reduced to selecting a pair of modal currents and evaluating the closed-form solution based solely on the associated eigenvalues. The characteristic modes \cite{HarringtonMautz_TheoryOfCharacteristicModesForConductingBodies, HarringtonMautz_ComputationOfCharacteristicModesForConductingBodies} are utilized here for their appealing properties, and the optimal composition of the characteristic modes is found as a particular outcome of the method paving the way to a productive workflow of unprecedented simplicity.

The problem to be solved is specified exactly in Section~\ref{Sec1_Formulation} and the roadmap of how the problem is tackled is provided.

\section{Formulation of the Optimization Problem}
\label{Sec1_Formulation}

The focal point of this paper is the minimization of quality factor $Q$ while preserving the self-resonance of the current at the given normalized frequency $ka$ (no external tuning elements are needed). Rigorously, this task can be expressed as
\begin{alignat}{3}
\label{OptTask1a}
& \text{minimize} && \text{quality factor } Q, \\
\label{OptTask1b}
&\text{subject to}\hspace{0.5cm} && \Wm - \We = 0,
\end{alignat}
in which the $\Wm$ and $\We$ represent magnetic and electric energy \cite{Harrington_TimeHarmonicElmagField}, respectively, and their vanishing difference indicates the resonance.

The problem (\ref{OptTask1a}), constrained by (\ref{OptTask1b}), is to be solved under the following assumptions:
\begin{itemize}
   \item small electrical size is considered, i.e. $ka < 1$,
   \item only perfectly conducting (PEC) bodies $\Omega$ are employed,
   \item only surface regions are treated.
\end{itemize}
With the first condition, the area is restricted to ESAs, though this is a minor technicality as these antennas suffer from high quality factor Q values, \cite{YaghjianBest_ImpedanceBandwidthAndQOfAntennas}, so the meaning of the optimization task (\ref{OptTask1a}) lies exactly within the range of $ka < 1$. The second condition means that no ohmic losses are  dealt with despite the fact quality factor $Q$ can always be reduced by adding ohmic losses, but this is not our intention. Notice that the ratio $G/Q$, where $G$ is the antenna gain, is neither dependent on the radiation nor on the ohmic losses at all, \cite{HarringtonMautz_ControlOfRadarScatteringByReactiveLoading}. The third condition intentionally excludes volumetric structures, which means that no materials are taken into account.

The optimization task (\ref{OptTask1a}) needs to be reformulated to become solvable. The procedure is as follows: the transition from the field to the source current quantities occurs in Section~\ref{S11_SourceConcept}, discretization of the continuous expressions into the matrix form is performed in Section~\ref{S12_MatrixRepreID}, the stored energy operator is formulated in Section~\ref{S13_MatrixRepreWsto}, the untuned and tuned quality factors $Q$ are introduced in Section~\ref{S14_Qfactor}, the optimization problem is reformulated to its solvable form in Section~\ref{S20_Reform}, then it is solved in Section~\ref{S21_decomp}. The solution in minimal representation is derived in Section~\ref{S21_CMdecomp} for a particular choice of the entire domain basis functions, and in Section~\ref{S3_Results} for an alternative basis. Results are provided in Section~\ref{S3_Results}. Various aspects of the proposed technique are addressed in the discussion in Section~\ref{S4_Discussion}.

\subsection{Source Concept Representation of the Antenna Parameters}
\label{S11_SourceConcept}

The antenna parameters are conventionally expressed as the functions of electric and magnetic fields, $\V{E}$ and $\V{H}$, respectively \cite{Balanis_Wiley_2005}. Instead of following this approach, we take advantage of the source concept, whereby within this concept, all electromagnetic quantities are represented as functions of source current density $\V{J}$. The source concept enables us to solve the problem in an efficient way, since we express and use all quantities exactly in the domain in which we would like to obtain the result -- the optimal currents. Among purely theoretical works \cite{Schwinger_SourcesAndElectrodynamics}, the source concept in classical electrodynamics has been pioneered by Harrington in the 1960s \cite{Harrington_MatrixMethodsForFieldProblems}, then extensively utilized after 2000 by Geyi \cite{Geyi_AMethodForTheEvaluationOfSmallAntennaQ}, Vandenbosch \cite{Vandenbosch_ReactiveEnergiesImpedanceAndQFactorOfRadiatingStructures}, Gustafsson et al. \cite{Gustafsson_OptimalAntennaCurrentsForQsuperdirectivityAndRP, GustafssonSohlKristensson_PhysicalLimitationsOfAntennasOfArbitraryShape_RoyalSoc, GustafssonFridenColombi_AntennaCurrentOptimizationForLossyMediaAWPL, GustafssonTayliEhrenborgCismasuNorbedo_MatlabCvxTutorial}.

The benefits of the source concept are obvious: the current is always of finite extent. Furthermore, the region of the non-zero current is electrically small for the ESAs. The current can be the subject of the structural \cite{Ethier_2012_substructure_TCM} or modal decomposition \cite{CapekHazdraEichler_AMethodForTheEvaluationOfRadiationQBasedOnModalApproach} if the smooth operators are discretized.

\subsection{Matrix Representation of the Integro-Differential Operators}
\label{S12_MatrixRepreID}

Demands on the operators' invertability and their spectral decomposition can rarely be met in the domain of continuous functions \cite{MorseFeshBach_MethodsOfTheoreticalPhysics}. For this reason, the entire region $\Omega$ of the non-zero current is discretized \mbox{$\Omega \rightarrow \OmegaN{N}$} into $N$ segments \cite{deLoeraRambauSantos_Triangulations, PetersonRayMittra_ComputationalMethodsForElectromagnetics}. It will be shown that the results are strongly dependent on the number of segments $N$ and their distribution. 

Consequently, the current density is represented as
\BE
\label{Discr3}
\V{J} \left(\V{r},\omega\right) \approx \sum\limits_{u=1}^U I_u\left(\omega\right) \V{f}_u \left(\V{r}\right)
\EE
in which \mbox{$\M{I} \in \mathbb{C}^{U \times 1}$} are the expansion coefficients,  $\V{f}_u$ are the basis functions \cite{PetersonRayMittra_ComputationalMethodsForElectromagnetics}, and all smooth operators are transformed according to
\BE
\label{Discr4}
\mathcal{L} \left(\V{J},\V{J}\right) \approx \Ivec^\herm \M{L} \Ivec, \quad \M{L} \in \mathbb{C}^{U \times U}
\EE
into a computationally feasible matrix representation. The time-harmonic quantities under the convention \mbox{$\V{J} \left(\V{r},t\right) = \RE\left\{\V{J}\left(\V{r},\omega\right) \mathrm{exp}\left(\J \omega t \right)\right\}$}, with $\omega$ being the angular freuquency, are used throughout the paper. The superscript $^\herm$ in (\ref{Discr4}) denotes the Hermitian transpose.

\subsection{Matrix Representation of the Stored Energy}
\label{S13_MatrixRepreWsto}

The stored energy functional for the smooth functions of the source current density has been derived in \cite{Vandenbosch_ReactiveEnergiesImpedanceAndQFactorOfRadiatingStructures}. The same results are published in \cite{CapekJelinekHazdraEichler_MeasurableQ, Gustaffson_StoredElectromagneticEnergy_PIER}, though by alternative approaches. The stored energy operator in matrix form was derived in \cite{CismasuGustafsson_FBWbySimpleFreuqSimulation, GustafssonFridenColombi_AntennaCurrentOptimizationForLossyMediaAWPL}, but it had already been anticipated by Harrington \cite{HarringtonMautz_ControlOfRadarScatteringByReactiveLoading}. The operator is defined by bilinear form
\BE
\label{StoredEnergy4}
\WSTO = \WM + \WE = \frac{1}{4\omega} \Ivec^\herm \XD \Ivec,
\EE
where
\BE
\label{StoredEnergy1}
\XD = \omega \frac{\partial \M{X}}{\partial \omega}
\EE
and in which $\M{X}$ is the reactance part of the impedance matrix $\M{Z}$, \mbox{$\M{Z} = \M{R} + \J \M{X}$}, \cite{Harrington_FieldComputationByMoM}.

According to the argumentation in \cite{Vandenbosch_ReactiveEnergiesImpedanceAndQFactorOfRadiatingStructures}, the stored energy (\ref{StoredEnergy4}) can also be separated into modified magnetic and electric energies which read
\begin{subequations}
\begin{align}
\label{StoredEnergy3A}
\WM &= \frac{1}{8\omega} \Ivec^\herm \XM \Ivec, \\
\label{StoredEnergy3B}
\WE &= \frac{1}{8\omega} \Ivec^\herm \XE \Ivec,
\end{align}
\end{subequations}
where
\begin{subequations}
\begin{align}
\label{StoredEnergy2A}
\XM &= \XD + \M{X}, \\
\label{StoredEnergy2B}
\XE &= \XD - \M{X},
\end{align}
\end{subequations}
respectively.

\subsection{Definition of Quality Factor Q}
\label{S14_Qfactor}

Quality factor~$Q$ is defined here in standard fashion \cite{IEEEStd_antennas}, with the exception that both the so-called untuned quality factor
\BE
\label{QualityFactor1}
\QU \left( \Ivec \right) = \frac{\omega\WSTO}{\Prad} = \frac{\Ivec^\herm \XD \Ivec}{2 \Ivec^\herm \M{R} \Ivec}
\EE
is defined, as well as the so-called tuned quality factor
\BE
\label{QualityFactor3}
\QT \left( \Ivec \right) = \QU \left( \Ivec \right) + \Qext \left( \Ivec \right) = \frac{\max \left\{\Ivec^\herm \XM \Ivec, \Ivec^\herm \XE \Ivec\right\}}{2 \Ivec^\herm \M{R} \Ivec}.
\EE
In both cases, radiated power \cite{Harrington_FieldComputationByMoM}
\BE
\label{QualityFactor2}
\Prad = \frac{1}{2} \Ivec^\herm \M{R} \Ivec
\EE
has been substituted. The amount of stored energy needed to tune quality factor $\QT$ to the resonance in (\ref{QualityFactor3}) can be determined from $\Qext$ which reads
\BE
\label{QualityFactor4}
\Qext \left( \Ivec \right) = \frac{\left|\Ivec^\herm \M{X} \Ivec\right|}{4\Ivec^\herm \M{R} \Ivec}.
\EE

\begin{figure}[t]
\begin{center}
  \includegraphics[width=\figwidth cm]{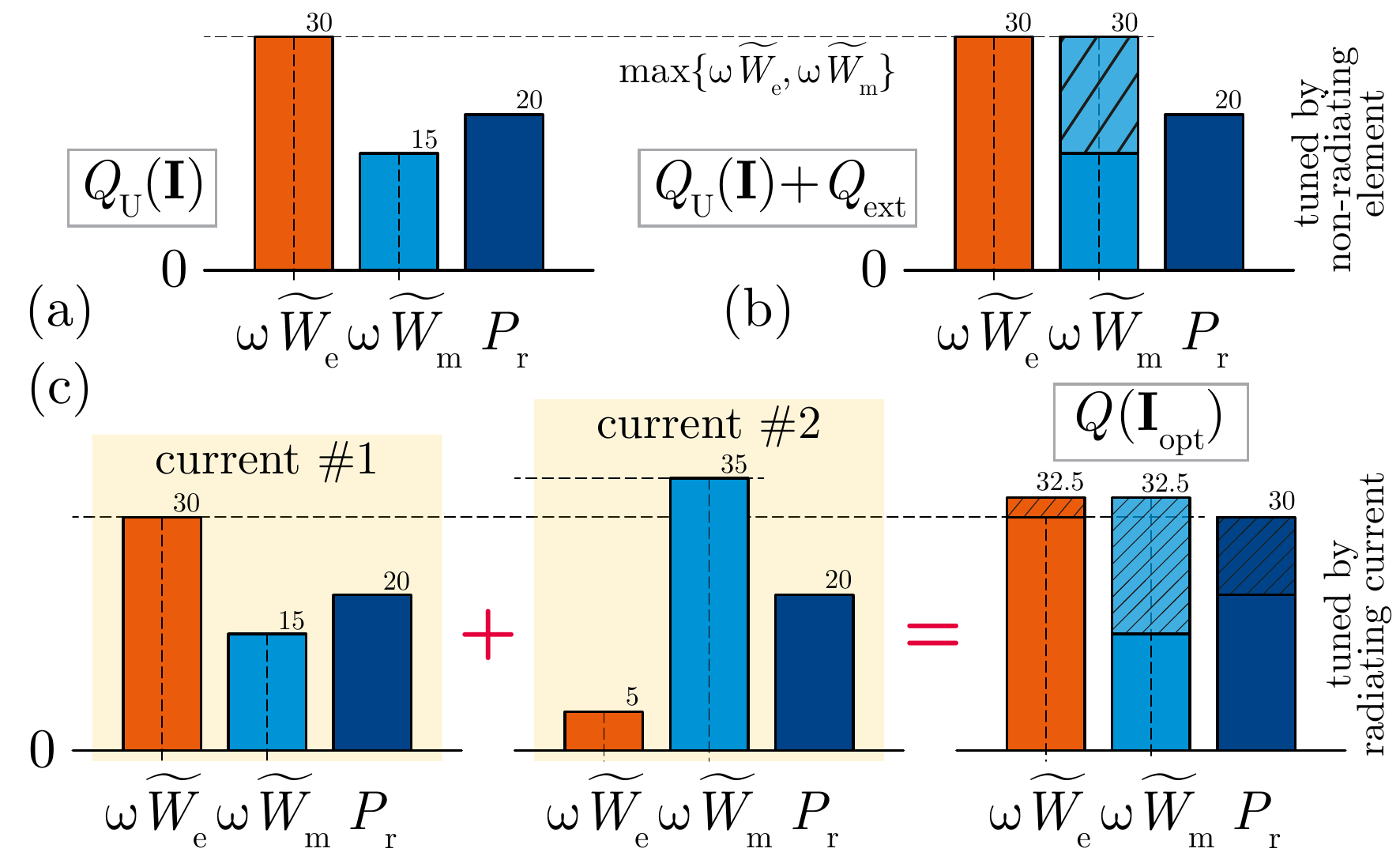}
  \caption{Demonstration of the principal difference between tuning by an external ideal (non-radiating) lumped element and tuning by a distributed current. The untuned quality factor $\QU$ is introduced in case a. The individual bars denote electric and magnetic energies multiplied by angular frequency and radiated power. To provide a numerical example, each bar is labelled by its height. Case b demonstrates conventional tuning by an external non-radiating element. Case c depicts the reduction of quality factor~$Q$ as a combination of two modal currents.}
  \label{Fig0}
\end{center}
\end{figure}

Now, we draw attention to the tuning procedure which is graphically depicted in Fig.~\ref{Fig0}. Let us suppose that current $\M{I}$ is not self-resonant. In such a case, the electric energy is not equal to the magnetic energy, \mbox{$\Qext \neq 0$}, see Fig.~\ref{Fig0}a. The first option is to tune the current by an external ideal lumped element, see Fig.~\ref{Fig0}b, which compensates for the lack of magnetic energy. Alternatively, the current can be tuned by an additional current as depicted in Fig.~\ref{Fig0}c. If the currents are properly chosen, the amount of stored energy is slightly higher than in case b, but the boost of the radiated power can decrease quality factor $Q$.

The tuning procedure involving the additional current is further elaborated as a key instrument in the minimization of quality factor~$Q$.

\section{Solution to the Optimization Problem}
\label{S2_Solution}

To solve the problem of the minimization of quality factor $Q$, all tools introduced above are employed in this section. The spectral decomposition of matrix operators $\M{R}$, $\M{X}$ and $\XD$ play the central role. The proper basis for unknown current $\Ivec$ will not be selected \textit{a priori}, but after a careful examination of the required basis' properties. 

\subsection{Reformulation of the Optimization Problem}
\label{S20_Reform}

The problem (\ref{OptTask1a})--(\ref{OptTask1b}) can now formally be rewritten as
\begin{alignat}{3}
\label{OptTask2a}
& \text{minimize} && \Ivec^\herm \XD \Ivec, \\
\label{OptTask2b}
& \text{subject to}\hspace{0.5cm} && \Ivec^\herm \M{X} \Ivec = 0, \\
\label{OptTask2c}
& && \Ivec^\herm \M{R} \Ivec = 1.
\end{alignat}
After careful inspection, the problem is truly recognized as non-convex \cite{BoydVandenberghe_ConvexOptimization}, which implies that it cannot be solved e.g., by convex optimization \cite{Gustafsson_OptimalAntennaCurrentsForQsuperdirectivityAndRP, GustafssonTayliEhrenborgCismasuNorbedo_MatlabCvxTutorial}, but -- surprisingly -- it can be solved directly if current $\Ivec$ is represented in an appropriate basis of the entire domain functions \cite{PetersonRayMittra_ComputationalMethodsForElectromagnetics}. 

\subsection{Modal Decomposition}
\label{S21_decomp}

Let us suppose that current $\Ivec$ is decomposed into modal currents $\Ivec_u$
\BE
\label{ModQ1}
\Ivec = \sum\limits_{u=1}^U \alpha_u \Ivec_u
\EE
with properties to be specified hereinafter. Substituting (\ref{ModQ1}) into (\ref{QualityFactor3}) yields
\BE
\label{ModQ2}
\QT \left(\Ivec\right) = \frac{\sum\limits_{v=1}^V \sum\limits_{u=1}^U \alpha_u^\herm \alpha_v \Ivec_u^\herm \XD \Ivec_v}{2 \sum\limits_{v=1}^V \sum\limits_{u=1}^U \alpha_u^\herm \alpha_v \Ivec_u^\herm \M{R} \Ivec_v} + \Qext \left(\Ivec\right).
\EE
The complexity of formulation (\ref{ModQ2}) can be substantially reduced if we choose $\Ivec_u$ to diagonalize the operator $\M{R}$, i.e., \mbox{$\Ivec_u^\herm \M{R} \Ivec_v/ 2 = \delta_{mn}$}, and we normalize the coefficients $\alpha_p$ as \mbox{$\alpha_p = \alpha_p / \alpha_1$}. Furthermore, it is assumed that modal currents $\Ivec_u$ can be chosen so as to satisfy
\BE
\label{OptTask3}
\Ivec_u^\herm \XD \Ivec_v = \delta_{uv},
\EE
and that only a combination of two modes 
\BE
\label{ModQ3}
\Ivec = \Ivec_1 + \alpha_2 \Ivec_2, \quad \left|\alpha_2 \right| \in \left[0,1\right],
\EE
guarantee the minimal quality factor~$Q$ defined as
\BE
\label{ModQ5}
\alpha_2 = \alpha_\iopt \Rightarrow \QT \left(\Ivec_\iopt\right) = \min\limits_\Ivec \left\{\QT \left(\Ivec\right)\right\}.
\EE
In (\ref{ModQ3}), the first mode, $\Ivec_1$, is denoted as the dominant mode and the second mode, $\Ivec_2$, is denoted as the tuning mode. The presumptions (\ref{OptTask3}) and (\ref{ModQ3}) simplify the general expression (\ref{ModQ2}) into the form
\BE
\label{ModQ4}
\QT \left(\Ivec\right) = \frac{\Ivec_1^\herm \XD \Ivec_1 + \left|\alpha_2\right|^2 \Ivec_2^\herm \XD \Ivec_2}{2 \left( 1 + \left|\alpha_2\right|^2 \right)} + \Qext \left(\Ivec\right).
\EE

\begin{figure}[t]
\begin{center}
  \includegraphics[width=\figwidth cm]{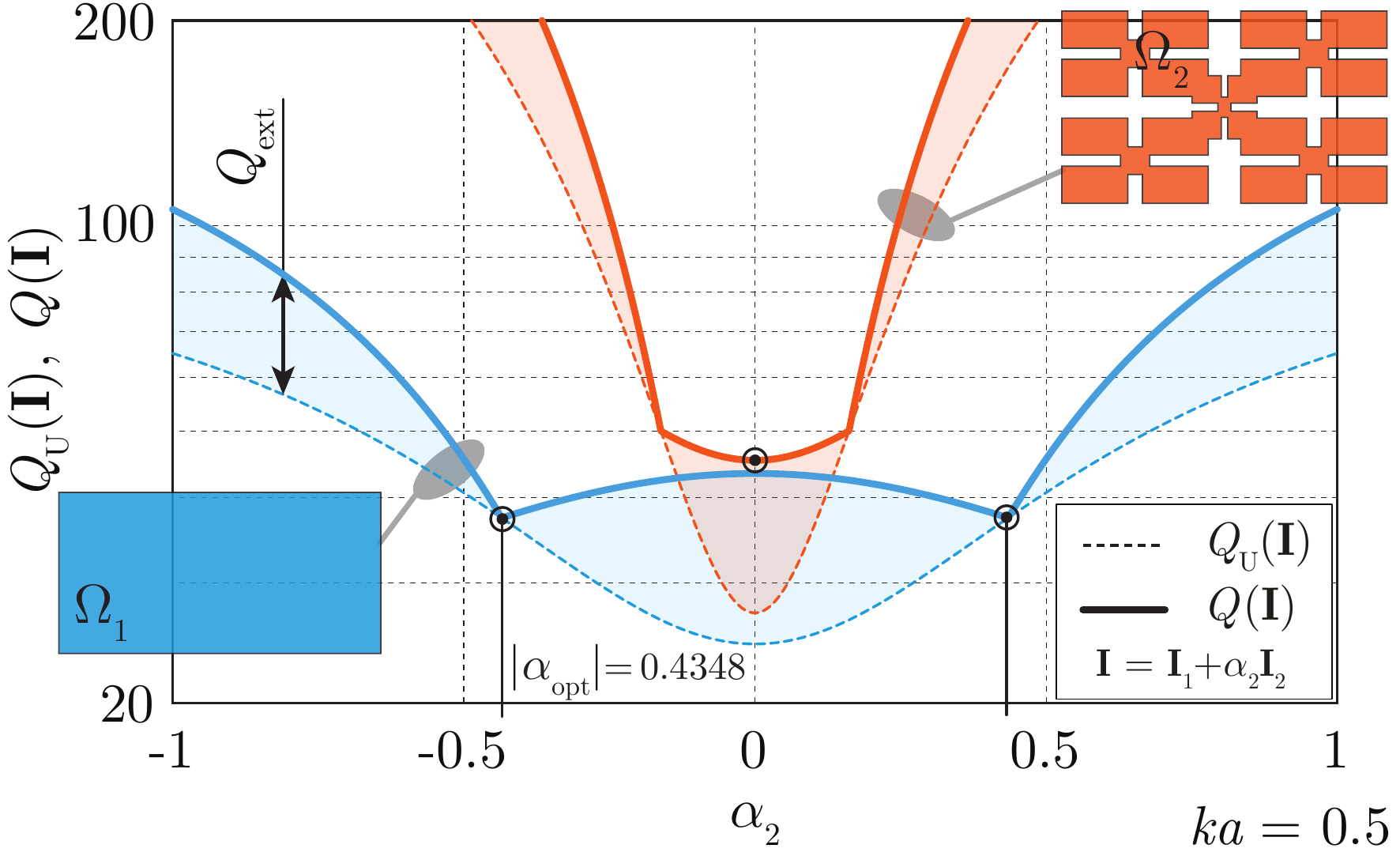}
  \caption{Untuned (dashed lines) and tuned (solid lines) quality factors for two structures of different geometrical complexity. The circular markers denote global extrema. The rectangular plate (blue, $\Omega_1$) is of dimensions $L \times L/2$, the fractal structure (orange, $\Omega_2$) has been produced according to Appendix~\ref{appB} with the particular choice of $p_2 = 0.2$, both structures were calculated at \mbox{$ka = 0.5$}.}
  \label{Fig1}
\end{center}
\end{figure}

The dependence of (\ref{ModQ4}) on $\alpha_2$ is briefly sketched out on an use-case of two shapes depicted in Fig.~\ref{Fig1}, the rectangular plate and the IFS fractal \cite{Falconer_FractalGeometry} of the second iteration. The rectangular plate can be easily tuned into its resonance by considering the proper combination of two modal currents, as seen by the solid blue curve in Fig.~\ref{Fig1}. Quality factor~$\QT\left(\Ivec\right)$ is significantly reduced compared to the quality factor of the dominant mode \mbox{$\QT\left(\Ivec_1\right)$}. This is in accordance with the example provided in Fig.~\ref{Fig0}c. On the other hand, quality factor~$Q$ of the fractal-like structure cannot be reduced by the superposition of two well-radiating modes and the tuning procedure needs to be done in a classical way by an external tuning element. This diverse behaviour is discussed in the following section where the condition for reduction of quality factor~$Q$ is provided.

The procedure how to obtain the proper basis and then select the dominant mode, the tuning modes and the coefficient $\alpha_2$ is a subject of the following sections, together with the general proof that the minimum quality factor~$Q$ is always realized by a self-resonant current.

\subsection{Characteristic Mode Decomposition}
\label{S21_CMdecomp}

An interesting basis which enables us to formulate (\ref{ModQ4}) is the characteristic mode basis \cite{HarringtonMautz_TheoryOfCharacteristicModesForConductingBodies} formulated as a generalized eigenvalue problem (GEP) \cite{Wilkinson_AlgebraicEigenvalueProblem}
\BE
\label{TCM1}
\M{X} \Ivec_u = \lambda_u \M{R} \Ivec_u.
\EE
The eigenvectors $\Ivec_u$ can be normalized as
\BE
\label{TCM2}
\frac{1}{2} \M{I}_u^\herm \M{Z} \M{I}_v = \left(1 + \J \lambda_u\right) \delta_{uv}
\EE
and the eigenvalues are equal to
\BE
\label{TCM3}
\lambda_u = \frac{\Ivec_u^\herm \M{X} \Ivec_u}{\Ivec_u^\herm \M{R} \Ivec_u}.
\EE
Unfortunately, the cross-terms (\ref{OptTask3}) are not identically equal to zero, but all results presented in Section~\ref{S3_Results} prove that the condition of no cross-terms is well aligned with the condition of $ka < 1$.

For our purposes, all characteristic modes $\Ivec_u$ are sorted according to the magnitude of their reactive power, $\left|\lambda_u\right|$, and labelled as inductive if \mbox{$\lambda_u > 0$}, and capacitive if \mbox{$\lambda_u < 0$} \cite{MartaEva_TheTCMRevisited}. This division will be utilized to combine modes in order to provide the resonance.

\begin{figure}[t]
\begin{center}
  \includegraphics[width=\figwidth cm]{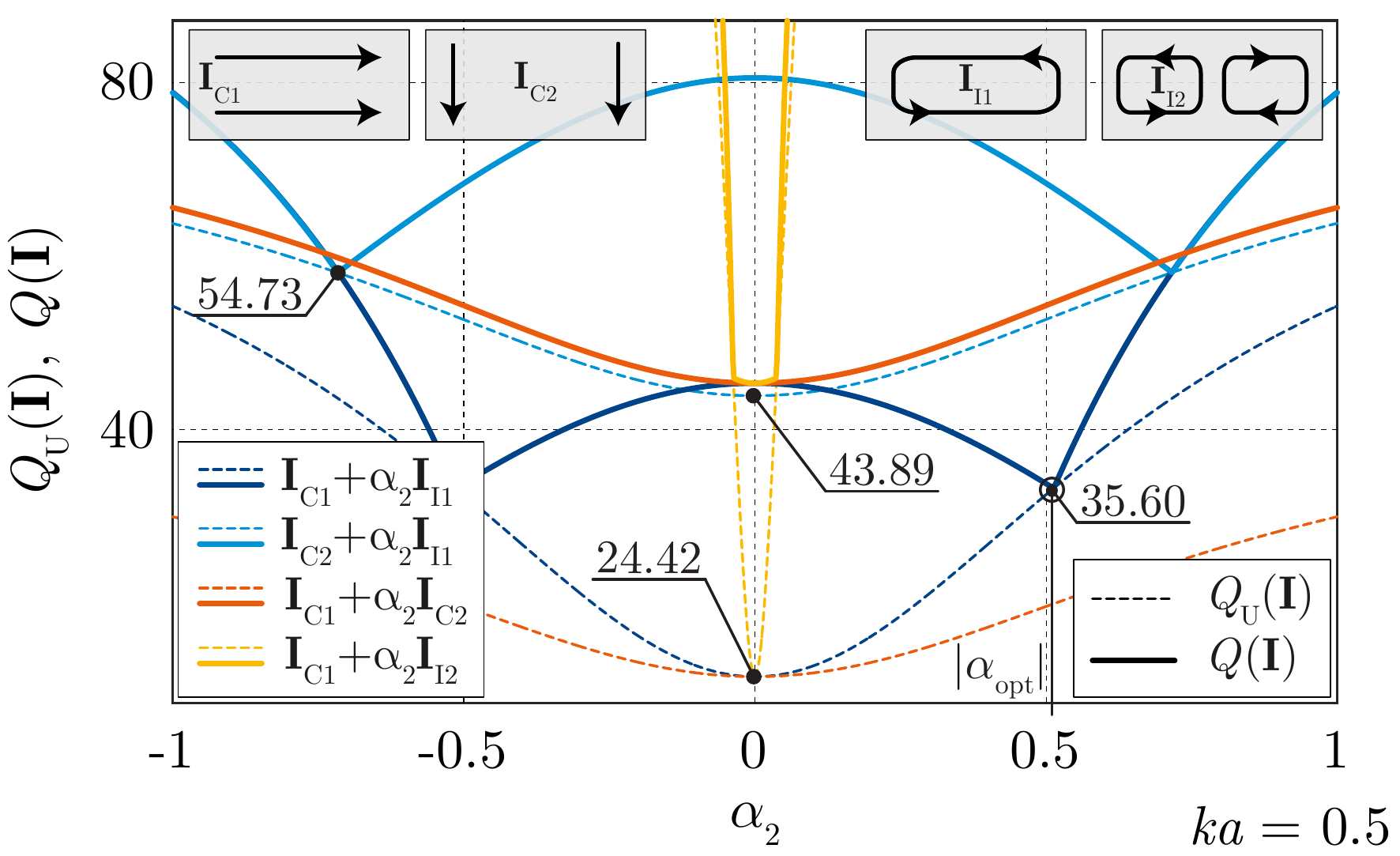}
  \caption{Demonstration of the technique reducing quality factor $Q$ by an additional tuning current. The untuned (dashed lines) and the tuned (solid lines) quality factors are depicted for different combinations of characteristic modes on a rectangular plate of dimensions $L\times L/2$.}
  \label{Fig2}
\end{center}
\end{figure}

Quality factor~$Q$ (\ref{ModQ4}) varies depending on which modes are combined. The impact of the selection of different characteristic modes is manifested in Fig.~\ref{Fig2}. Two capacitive (C1, C2) and two inductive (I1, I2) modes are combined two by two. As expected from Fig.~\ref{Fig1}, the minimal quality factor~$Q$ ($\QT = 35.60$) is obtained for a self-resonant current (no external tuning is necessary), which is obtained as a combination of the dominant mode ($\Ivec_\mathrm{C1}$) with the inductive tuning mode ($\Ivec_\mathrm{I1}$). The global minimum is highlighted by a circle marker. A combination of two dominantly electric modal currents ($\Ivec_\mathrm{C1}$, $\Ivec_\mathrm{C2}$), or two magnetic modal currents, cannot be tuned to the resonance without an external tuning element.

In order to combine the modes optimally, the mode with the lowest untuned quality factor~$\QU$ (\ref{QualityFactor1}) is selected as the dominant mode $\Ivec_1$. Since the electric energy predominates the magnetic energy in the ESA region \cite{GustafssonSohlKristensson_PhysicalLimitationsOfAntennasOfArbitraryShape_RoyalSoc, Yaghjian_MinimumQforLossyAndLosslessESA}, its behaviour is generally capacitive. The tuning mode should be selected so that the quality factor~$Q$ is minimized. While this mode cannot be selected \textit{a priori} (since it depends on the shape), all numerical results confirm that a good candidate is the first inductive mode. 

Having now a pair of modes, the coefficient $\alpha_\iopt$ can be found. It is proved in Appendix~\ref{appCM1}, and confirmed by all numerical examples presented in Section~\ref{S3_Results}, that the optimal composition of modal currents guarantees the quality factor $\Qext \left(\Ivec\right)$ is zero, i.e.
\BE
\label{TCM6}
\alpha_v = \alpha_\iopt \Rightarrow \Qext \left(\Ivec_\iopt\right) = 0.
\EE
Using this formula together with (\ref{QualityFactor4}) and (\ref{TCM3}) we have
\BE
\label{TCM7}
\Qext \left(\Ivec\right) = \frac{\left|\lambda_1 + \left|\alpha_\iopt\right|^2 \lambda_2 \right|}{4 \left(1 + \left|\alpha_\iopt\right|^2 \right)} = 0,
\EE
which gives
\BE
\label{TCM9}
\left| \alpha_\iopt\right|^2 = -\frac{\lambda_1}{\lambda_2},
\EE
and finally
\BE
\label{TCM10}
\alpha_\iopt = \sqrt{-\frac{\lambda_1}{\lambda_2}} \mathrm{e}^{\J \varphi},\quad \varphi\in\left[-\pi,\pi\right], \quad \lambda_2 \neq 0.
\EE
It can be deduced from (\ref{TCM9}) that the tuning mode has to be of opposite capacitive-inductive behaviour than the dominant mode (usually inductive). The condition (\ref{ModQ3}) also dictates that $0 \leq \left|\lambda_1 \right| \leq \left|\lambda_2 \right|$.  Notice also that the phase between the modes $\varphi$ can be chosen arbitrarily. 

Thanks to the knowledge of optimal ratio $\alpha_\iopt$, we can determine which modal combination yields a smaller quality factor~$\QT$ than the dominant mode and which does not, see two dissimilar examples in Fig.~\ref{Fig1}. The condition is derived in Appendix~\ref{appCM2} and it reads
\BE
\label{TCMopt1}
\QT\left(\Ivec\right) < \QT\left(\Ivec_1\right) \Leftrightarrow \Ivec_v^\herm \XD \Ivec_2 - \Ivec_1^\herm \XD \Ivec_1 < \left|\lambda_1 \right| + \left|\lambda_2 \right|.
\EE

The final step is to express the minimal quality factor~$Q$ which can be done by substituting (\ref{TCM6}) and (\ref{QualityFactor1}) into (\ref{ModQ4}) which yields
\BE
\label{TCM11}
\QT \left(\Ivec_\iopt\right) = \frac{\QU \left(\Ivec_1\right) + \left|\alpha_\iopt\right|^2 \QU\left(\Ivec_2\right)}{1 + \left|\alpha_\iopt\right|^2}.
\EE
A great advantage of the formula (\ref{TCM11}) is that only two modes with small eigenvalues (typically dominant mode and inductive mode with smallest eigenvalue) are needed and, thus, the notoriously known issue with the ill-conditioned weighting part of the matrix pencil in (\ref{TCM1}), $\M{R} \nsucc \M{0}$, is not challenged \cite{CapekHazdraEichler_ComplexPowerRatioFunctionalForRadiatingStructures}.

\subsection{Problem Redefinition in Alternative Basis}
\label{S22_XMdecomp}

Alternative basis
\BE
\label{XM1}
\XD \Ivec_p = q_p \M{R} \Ivec_p,
\EE
with the eigenvalues
\BE
\label{XM2}
q_p = \frac{\Ivec_p^\herm \XD \Ivec_p}{\Ivec_p^\herm \M{R} \Ivec_p}
\EE
is utilized in this section so that (\ref{OptTask3}) is satisfied.

In comparison with (\ref{QualityFactor1}), (\ref{XM1}) provides the formula
\BE
\label{XM3}
q_p = 2 \QU \left( \Ivec_p \right).
\EE
To differentiate between the characteristic modes (\ref{TCM1}) and the modes originated in (\ref{XM1}), the different subindices $u$, $v$ and $p$, $q$ are thoroughly used. The radiated power is once more normalized
\BE
\label{XM4}
\frac{1}{2} \Ivec_p^\herm \M{R} \Ivec_p = 1
\EE
and another simplification appears from the fact that the modes are orthogonal with respect to the $\XD$ operator. Consequently, the formula (\ref{ModQ2}) is reduced to a convenient form
\BE
\label{XM5}
\QT \left(\Ivec\right) = \frac{\sum\limits_p \left|\alpha_p\right|^2 \QU \left(\Ivec_p\right)}{\sum\limits_p \left|\alpha_p\right|^2} + \Qext \left(\Ivec\right),
\EE
where (\ref{XM2})--(\ref{XM4}) have been substituted. The minimization of (\ref{XM4}) is tightly connected with the tuning procedure and follows the same line of reasoning as in Section~\ref{S21_CMdecomp}. The tuning ratio between the dominant mode $\Ivec_p$, with the lowest $\QU$, and the suitable tuning mode $\Ivec_q$ is
\BE
\label{XM6}
\alpha_\iopt = \sqrt{\frac{\Ivec_p^\herm \M{X} \Ivec_p}{\Ivec_q^\herm \M{X} \Ivec_q}} \mathrm{e}^{\J \varphi},\quad \varphi\in\left[-\pi,\pi\right], \quad \Ivec_q^\herm \M{X} \Ivec_q \neq 0,
\EE
which leads to exactly the same expression as (\ref{TCM11}). 

The decomposition (\ref{XM1}) constitutes an interesting alternative to characteristic modes since the cross-terms (\ref{OptTask3}) are identically zero. Notice that both decompositions are mentioned in the context of quality factor~$Q$ in \cite{HarringtonMautz_ControlOfRadarScatteringByReactiveLoading, GustafssonTayliEhrenborgCismasuNorbedo_MatlabCvxTutorial} as well, however, the tuning procedure and the minimization are not performed.

\section{Results}
\label{S3_Results}
The results presented in this section have been acquired in the basis of the characteristic modes (\ref{TCM1}). It will be proved that for electrically small dimensions, the optimal current, with respect to the minimal quality factor~$Q$, can always be found this way. In other words, the non-zero cross-terms (\ref{OptTask3}) can be safely neglected, thus the existent codes \cite{atom, feko} can be easily used. If required, especially for electrically larger structures, the alternative decomposition (\ref{XM1}) should be used to prevent the influence of the energy cross-terms. Before we turn to an inspection of the results, the particular implementation details are specified. 

\subsection{Implementation Details}
\label{S30_Implementation}

All dimensions are normalized to $ka$ in which $k$ is the wavenumber and $a$ is a radius of a sphere completely surrounding the sources. The smooth surfaces $\Omega$ have been triangularized and the density of the mesh grid has been normalized with respect to the original object's area and the surface of a sphere of radius $a$ as
\BE
\label{Ndiscr1}
\Nmeas \left(\Omega\right) = \frac{4 \pi a^2 N}{\int\limits_\Omega \D{S}},
\EE
in which $N$ is number of triangles. At least five basis functions have been systematically used per the narrowest part of the structure which ensured that the inductive modes could appear. The quality of the triangles\footnote{The quality $\kappa_i$ of a triangle $T_i$ is defined as \mbox{$\kappa_i = 4 \sqrt{3} A_i / \left(h_{i,1}^2+h_{i,2}^2+h_{i,3}^2\right)$}, in which $A_i$ is the area and $h_{1,2,3}$ are the lengths of the sides of the $i$th triangle.} were at least $0.5$.

The RWG basis functions \cite{RaoWiltonGlisson_ElectromagneticScatteringBySurfacesOfArbitraryShape} have been utilized to transform the optimization problem into matrix form. The terms $\left[Z_{mn}\right]$ of the impedance matrix can be calculated according to \cite{RaoWiltonGlisson_ElectromagneticScatteringBySurfacesOfArbitraryShape}. Alternatively the rooftop functions \cite{GlissonWilton_SimpleAndEfficientNumericalMethodsForProblemsOfEMRad} can be utilized as well. The impedance matrices have been acquired in the AToM package \cite{atom}, the GEP (\ref{TCM1}) has been solved by the iteratively restarted Arnoldi (IA) method \cite{Saad_NumericalMethodsForLargeEigenvalueProblems} and the postprocessing has been done in Matlab \cite{matlab}. 

To normalize the values of quality factor~$Q$, the fundamental bounds for a spherical shell of radius $a$ are introduced for a single (TM) mode operation \cite{Chu_PhysicalLimitationsOfOmniDirectAntennas}
\BE
\label{QchuRes1}
\QchuTM = \frac{1}{\left(ka\right)^3} + \frac{1}{ka}
\EE
and double (combination of TM and TE) mode operation \cite{McLean_AReExaminationOfTheFundamentalLimitsOnTheRadiationQofESA}
\BE
\label{QchuRes2}
\QchuTMTE = \frac{1}{2} \left(\frac{1}{\left(ka\right)^3} + \frac{2}{ka} \right).
\EE
The fundamental bound (\ref{QchuRes2}) is attainable only if the electric and the magnetic currents are combined.

\subsection{Quality Factor of a Rectangular Plate}
\label{S31_RectPlate}

\begin{figure}[t]
\begin{center}
  \includegraphics[width=\figwidth cm]{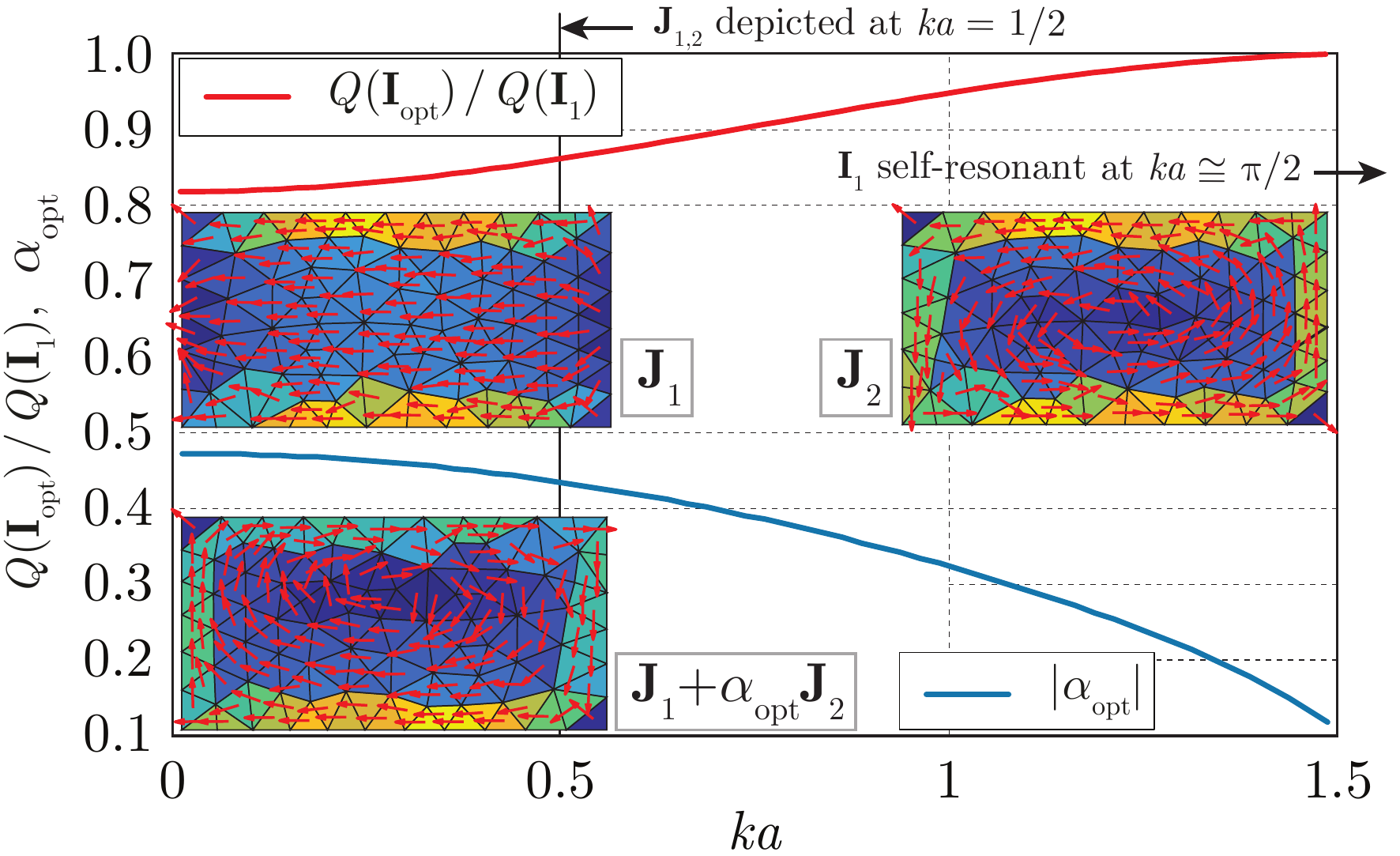}
  \caption{The reduction of quality factor $\QT$ by a proper combination of characteristic modes on a rectangular plate. The same structure as in Fig.~\ref{Fig2} has been used with discretization $\Nmeas = 2165$. For better visibility, the resulting current densities are shown on a much coarser discretization. The reduction rate is depicted as a ratio to tuned quality factor of the dominant characteristic mode (red solid line). The tuning coefficient $\alpha_\iopt$ (blue curve) has been calculated according (\ref{TCM6}), $\varphi = 0$, and verified by sweeping $\alpha_2$ between -1 and 1 in 1001 points. The resulting current density \mbox{$\M{J}_\iopt = \M{J}_1 + \alpha_\iopt \M{J}_2$} is depicted together with the contributing modal currents $\M{J}_1$ and $\M{J}_2$ for \mbox{$ka = 1/2$}.}
  \label{Fig3}
\end{center}
\end{figure}

The minimal quality factor~$Q$ for a rectangular plate of dimensions $L\times L/2$ of the electrical size $ka = 0.5$ is depicted in Fig.~\ref{Fig1} as a function of tuning coefficient $\alpha_2$. The minimum occurs for $\alpha_\iopt \approx 0.4848$, where the untuned quality factor $\QU$ is equal to quality factor $\QT$. This confirms that the minimal value occurs for the self-resonant current. The curves representing quality factors $\QU$ and $\QT$ are symmetric with respect to $\alpha_2$ which reveals\footnote{Using \mbox{$\Ivec = \Ivec_1 + \alpha_2 \Ivec_2$} in the denominator of (\ref{QualityFactor1}), we get \mbox{$\Ivec^\herm \M{X} \Ivec = \Ivec_1^\herm \M{X} \Ivec_1 + 2 \RE\left\{\alpha_2\right\}\Ivec_1^\herm \M{X} \Ivec_2 + \left|\alpha_2\right|^2\Ivec_2^\herm \M{X} \Ivec_2$}. Since \mbox{$\alpha_2 \in \left[-1,1\right] $}, the only possibility for quality factor \mbox{$\QU\left(\Ivec\right)$} being symmetric with respect to the axis \mbox{$\alpha_2 = 0$} is \mbox{$\Ivec_1^\herm \M{X} \Ivec_2 = 0$}.} that the cross-terms can be neglected. This implies that only the $\left|\alpha_\iopt\right|^2$ value is relevant for the minimal quality factor~$Q$. It means that the dominant mode and the tuning modes, which are depicted in Fig.~\ref{Fig3} as $\M{J}_1$ and $\M{J}_2$, can be shifted by arbitrary phase. Clearly, all possible solutions lie on the circle in the complex $\alpha_2$-plane with radius \mbox{$\sqrt{\left|\lambda_1 / \lambda_2 \right|}$} and centre at \mbox{$\alpha_2 = 0$}. Quality factor~$\QT$ of various selections of the dominant and the tuning mode is depicted in Fig.~\ref{Fig2}. It reveals that there is a unique combination of two modes yielding the lowest quality factor~$Q$.

A comparison of the externally tuned quality factor of the dominant mode \mbox{$\QT\left(\Ivec_1\right)$} and the optimal combination of two modes \mbox{$\QT\left(\Ivec_\iopt\right)$} is depicted in Fig.~\ref{Fig3} as a function of $ka$. It can be seen that for \mbox{$ka\rightarrow 0$}, the reduction is close to $80\,\%$ of quality factor~$Q$ of the dominant mode. As the dominant mode approaches its natural resonance (approx. at $\pi /2$), the contribution of the tuning mode starts to approach zero (see blue curve in Fig.~\ref{Fig3}). The resulting current density is depicted, together with the original modal contributions in Fig.~\ref{Fig3}, for \mbox{$ka = 0.5$}.

\subsection{Quality Factor of a Spherical Shell}
\label{S32_SphShell}

\begin{figure}[t]
\begin{center}
  \includegraphics[width=\figwidth cm]{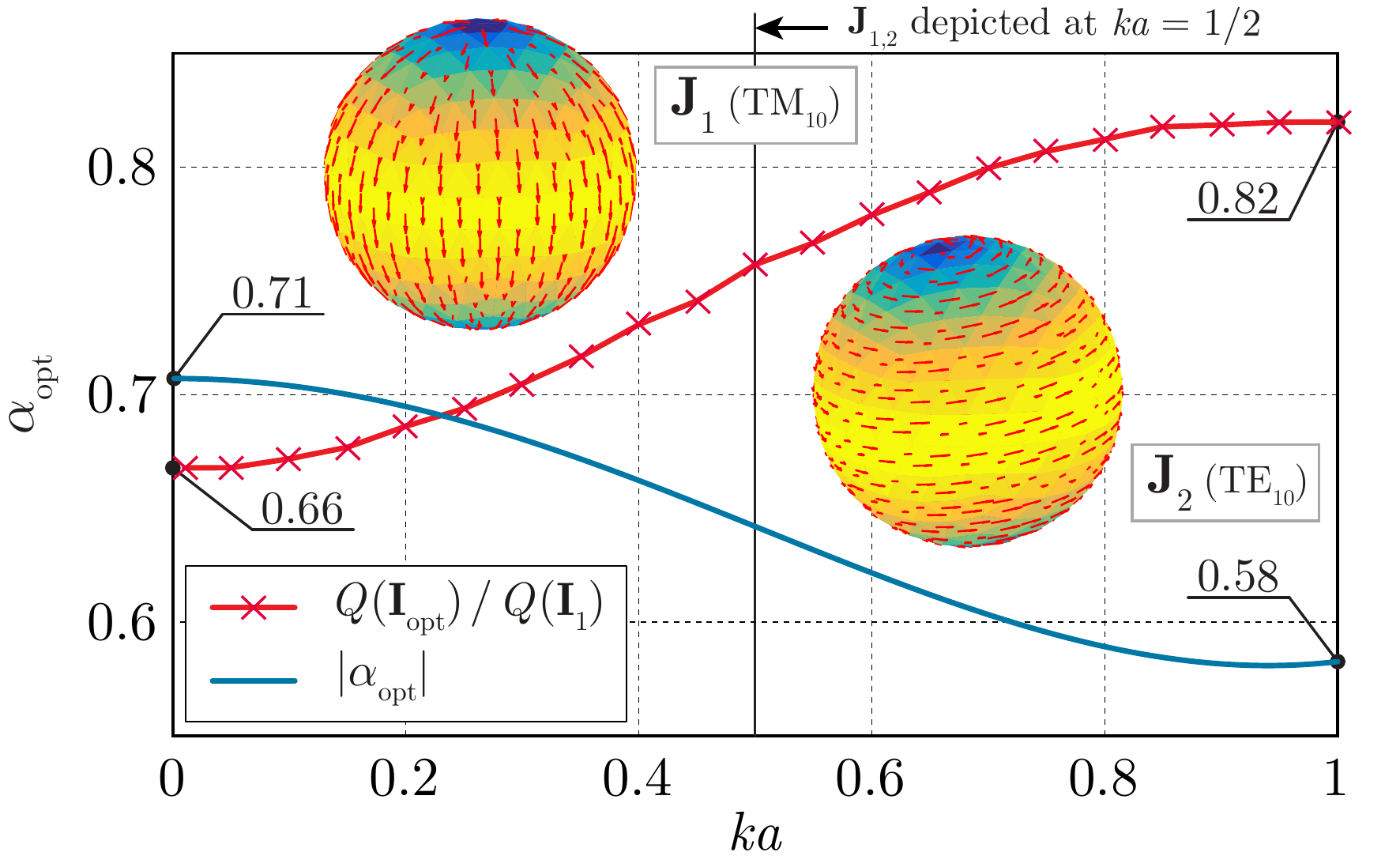}
  \caption{The reduction of quality factor $\QT$ by a proper combination of the characteristic modes on a spherical shell of radius $a$. The meaning of the curves is the same as in Fig.~\ref{Fig3}. The optimal ratio between modes (blue line) has been calculated analytically according to (\ref{ExSph1}). Red cross markers depict the $ka$ value on which the data have been calculated. The spherical shell has been discretized with a ratio of $\Nmeas = 506.1$. The current densities of two contributing characteristic modes $\M{J}_1$ and $\M{J}_2$ are depicted at $ka = 1/2$.}
  \label{Fig4}
\end{center}
\end{figure}

A spherical shell is a canonical shape of well-explored electromagnetic behaviour. For this reason, it is carefully studied in this section. The same study performed for the rectangular shape in Fig.~\ref{Fig3} is used in Fig.~\ref{Fig4} for a spherical shell of a radius $a$. To obtain the minimum quality factor~$Q$, the dominant spherical $\mathrm{TM}_{10}$ and $\mathrm{TE}_{10}$ modes are combined, see corresponding current densities $\M{J}_1$ and $\M{J}_2$ for \mbox{$ka=0.5$} in Fig.~\ref{Fig4}. The overall tuned quality factor~$Q$ can significantly be reduced, up to $66\,\%$ of the quality factor~$Q$ of the $\mathrm{TM}_{10}$.

A spherical shell provides us with the interesting opportunity to calculate the optimal tuning coefficient $\alpha_\iopt$ analytically. For a sphere, the spherical harmonics coincide with characteristic modes \cite{Garbacz_TCMdissertation}, which make it possible to evaluate (\ref{TCM9}) analytically from (\ref{TCM3}) since the radiated and the reactive power is known. See Appendix~\ref{appA} for its derivation. The resulting coefficient reads
\BE
\label{ExSph1}
\alpha_\iopt = \sqrt{\left| \displaystyle\frac{1 - ka \displaystyle\frac{\YBSL{0}{ka}}{\YBSL{1}{ka}}}{1 - ka \displaystyle\frac{\JBSL{0}{ka}}{\JBSL{1}{ka}}} \right|} \mathrm{e}^{\J \varphi}
\EE
where $\JBSL{n}{ka}$ and $\YBSL{n}{ka}$ are spherical Bessel functions of the $n$th order and of the first and the second kind, respectively. The optimal coefficient (\ref{ExSph1}) is valid for $ka$ under the resonance of the dominant $\mathrm{TM}_{10}$ and $\mathrm{TE}_{10}$ modes. The tuning coefficient varies from \mbox{$\left|\alpha_\iopt\right|^2 = 1/2$} for $ka \rightarrow 0$ to \mbox{$\left|\alpha_\iopt\right|^2 \approx 0.339$} for $ka = 1$ which means that the tuning mode is significant even for relatively high values of $ka$. 

\begin{figure}[t]
\begin{center}
  \includegraphics[width=\figwidth cm]{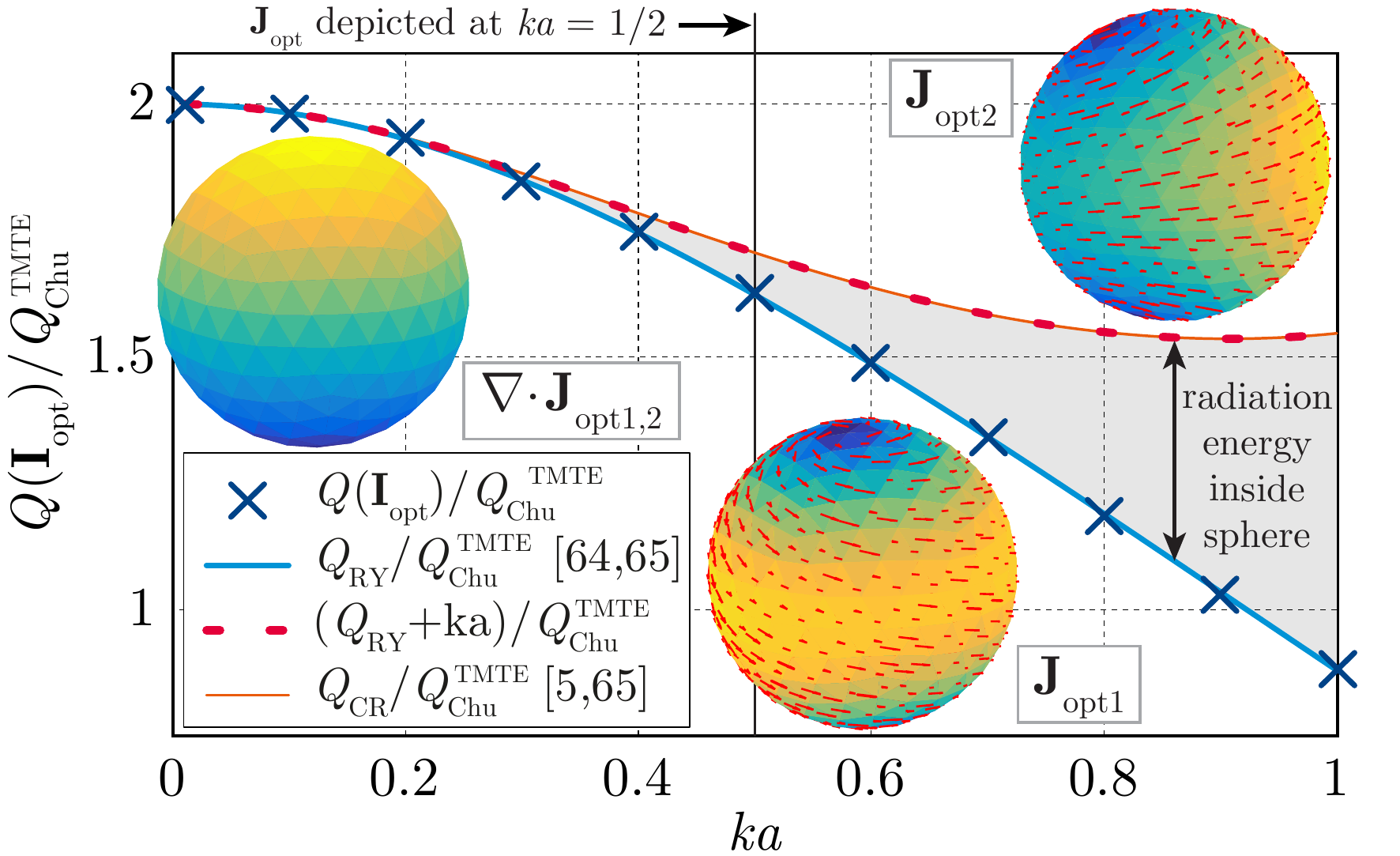}
  \caption{The tuned quality factor $\QT \left(\Ivec_\iopt\right)$ for a spherical shell of radius $a$ (cross markers) is normalized to the Chu's lower bound $\QchuTMTE$ for a combination of spherical $\mathrm{TM}_{10}$ and $\mathrm{TE}_{10}$ modes (\ref{QchuRes2}) and depicted as a function of $ka$. The optimal current composition $\Ivec_\iopt$, found in Fig.~\ref{Fig4}, has been used. The resulting ratio is compared with the analytically calculated quality factors $Q_\mathrm{RY}$ and $Q_\mathrm{CR}$, proposed in \cite{Rhodes_ObservableStoredEnergiesOfElectromagneticSystems} and \cite{CollinRotchild_EvaluationOfAntennaQ}, respectively, with optimal composition of the electric currents \cite{CapekJelinekHazdraEichler_QofSphere} substituted. The stored energy formulations used in quality factors $Q_\mathrm{RY}$ and $Q_\mathrm{CR}$ vary the radiation energy extraction, leading to a difference of $ka$. Using the same extraction, the results are in perfect agreement. Two resulting current densities are depicted for \mbox{$ka = 0.5$}, including their charge density.}
  \label{Fig5}
\end{center}
\end{figure}

The minimum quality factor~$\QT \left(\Ivec_\iopt\right)$ resulting from (\ref{ExSph1}) is depicted in Fig.~\ref{Fig5} for 11~equidistantly spaced points (cross marks). The values are in excellent numerical agreement with quality factor $Q_\mathrm{RY}$ which stands for a quality factor~$Q$ based on the stored energy calculated from the field formulation proposed by Rhodes \cite{Rhodes_ObservableStoredEnergiesOfElectromagneticSystems} and refined by Yaghjian and Best \cite{YaghjianBest_ImpedanceBandwidthAndQOfAntennas} with optimal composition for TM and TE modes found in \cite{CapekJelinekHazdraEichler_QofSphere}. Within this scheme, and similarly as in this paper \cite{Vandenbosch_ReactiveEnergiesImpedanceAndQFactorOfRadiatingStructures}, the radiating energy is subtracted everywhere \cite{CapekJelinek_VariousInterpretationOfTheStoredAndTheRadiatedEnergyDensity}, including the interior of the sphere. This can be fixed by the $ka$ correction \cite{Gustaffson_StoredElectromagneticEnergy_PIER} (see dashed line in Fig.~\ref{Fig5}), so we have excellent agreement with $Q_\mathrm{CR}$, as calculated according Collin and Rothschild's extraction \cite{CollinRotchild_EvaluationOfAntennaQ} for the same modal composition.

As seen in Fig.~\ref{Fig5}, the fundamental bound (\ref{QchuRes2}) cannot be reached by purely electrical currents (even though they generate both TM and TE modes) since a portion of energy is stored inside the sphere. Combining the proposed technique with the magnetic currents $\M{M}$ \cite{Kim_LowerBoundsOnQForFinizeSizeAntennasOfArbitraryShape}, the fundamental bound can be reached.

The current densities $\M{J}_{\iopt 1}$ and $\M{J}_{\iopt 2}$, which minimize quality factor~$Q$, are depicted in Fig.~\ref{Fig5}. Generally, six degenerated solutions can be found. Both current densities have the same charge distributions, see the top-left part of Fig.~\ref{Fig5}. It should be mentioned that one of the solutions (bottom middle) resembles Best's helix antenna \cite{Best_LowQelectricallySmallLinearAndEllipticalPolarizedSphericalDipoleAntennas} which can be manufactured by adding slots and a feeding network at the equator of the sphere. However, the second solution (top right) has a counter-intuitive distribution of the current density. After careful inspection, one can realize that this solution is, in fact, quite similar to the optimal current density found on a rectangle plate with maximum located asymmetrically at one of the margins, see Fig.~\ref{Fig3}. 

Notice, that the phase between the modes is not relevant for the minimum quality factor~$Q$ as the optimal composition depends only on $\left|\alpha_\iopt\right|^2$ squared, but it will be shown that the phase between $\mathrm{TM}_{10}$ and $\mathrm{TE}_{10}$ modes plays a major role in maximizing the $G/Q$ ratio.

\subsection{Study of the $G/Q$ Ratio}
\label{S33_DverQ}

This work is not aimed at the optimization of the $G/Q$ ratio, but a by-product of quality factor~$Q$ minimization is the following observation
\BE
\label{DoverQ1}
\frac{G_{\tau} \left(\Ivec_\iopt\right)}{\QT \left(\Ivec_\iopt\right)} \approx \max\limits_\Ivec \left\{ \frac{G_{\tau} \left(\Ivec\right)}{\QT \left(\Ivec\right)}\right\},
\EE
in which $\tau$ denotes the polarization. The expression (\ref{DoverQ1}) holds for $ka < 1$ very well and tells us that the current, which is optimal with respect to the minimal quality factor~$Q$, also (at least approximately) maximizes the $G/Q$ ratio. To justify this statement, the $G/Q$ ratio is rewritten as
\BE
\label{DoverQ12}
\frac{G_{\tau} \left(\Ivec_\iopt\right)}{\QT \left(\Ivec_\iopt\right)} = \frac{\Ivec_\iopt^\herm \M{D}_\tau \Ivec_\iopt}{\QT \left(\Ivec_\iopt\right)} = \frac{D_{\tau} \left(\Ivec_\iopt\right)}{\QT \left(\Ivec_\iopt\right)},
\EE
in which the directivity operator $\M{D}_\tau$ for a particular choice of \mbox{$\tau = \left\{\theta,\phi\right\}$} or \mbox{$\tau = \left\{x,y,z\right\}$} is derived in Appendix~\ref{appC}. 

\begin{figure}[t]
\begin{center}
  \includegraphics[width=\figwidth cm]{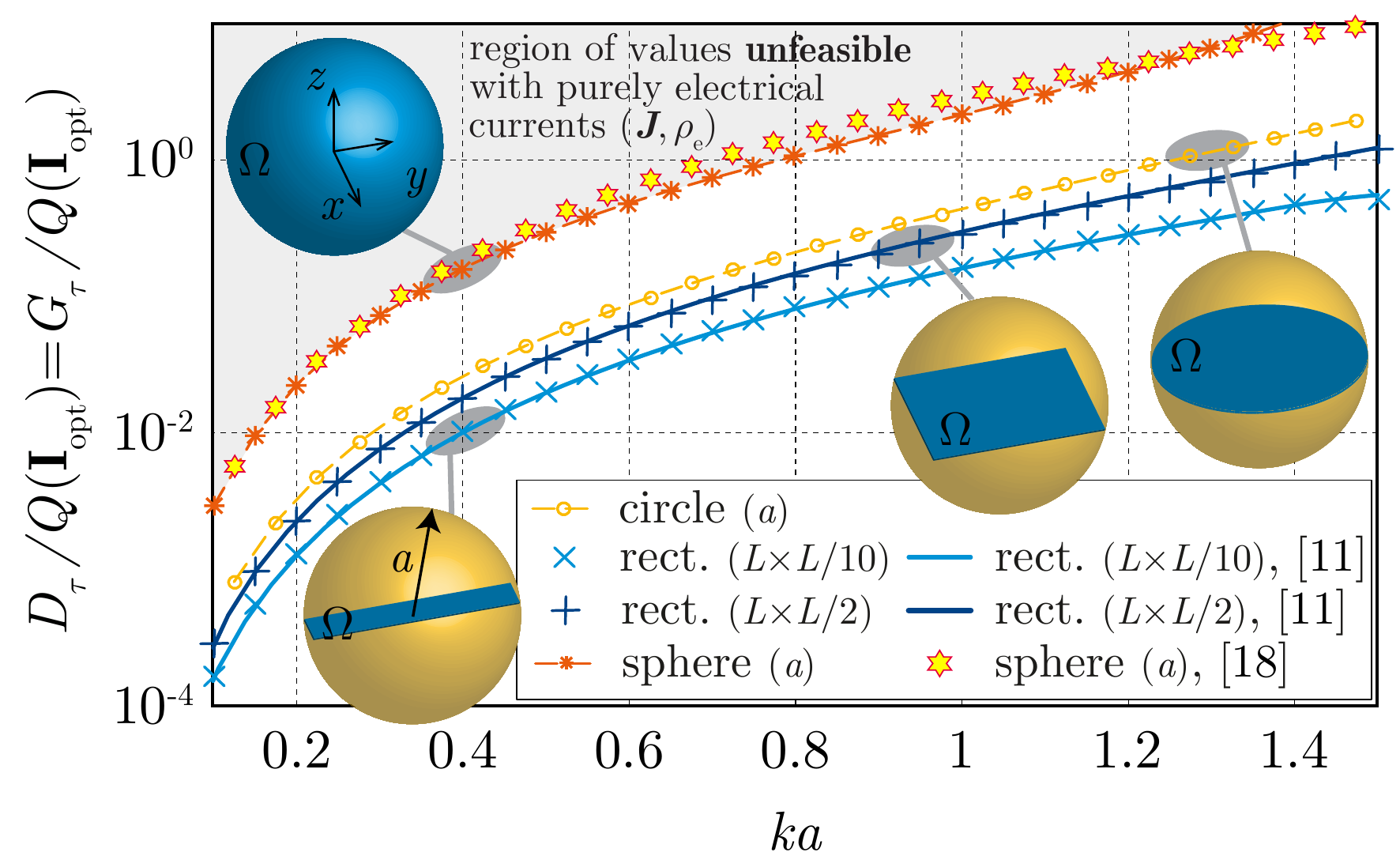} 
  \caption{The comparison of $G_\tau / \QT\left(\Ivec_\iopt\right)$ for a variety of simple connected shapes. The results for a rectangle are compared with the bounds found in \cite{GustafssonTayliEhrenborgCismasuNorbedo_MatlabCvxTutorial} with $\tau = y$ calculated in the $z$-direction. In the case of the sphere, the resulting current has been postprocessed so that the maximal partial directivity, \mbox{$\tau = \theta$}, points to the $x$-axis, see also Fig.~\ref{Fig10}. Since two different current densities minimize quality factor~$Q$, see Fig.~\ref{Fig5}, the solution with higher partial directivity ($\M{J}_{\iopt 2}$) has been compared with the bounds analytically derived from the static polarizability for an electric-magnetic combination in \cite{GustafssonSohlKristensson_PhysicalLimitationsOfAntennasOfArbitraryShape_RoyalSoc}. The grey-filled region is unfeasible by any electric current.}
  \label{Fig7}
\end{center}
\end{figure}
The validity of (\ref{DoverQ1}) is studied in Fig.~\ref{Fig7} for a dipole, a rectangular plate, a circular plate and a spherical shell. The dipole and the rectangular plate have the same dimensions as in \cite{GustafssonTayliEhrenborgCismasuNorbedo_MatlabCvxTutorial}, in which the RHS of (\ref{DoverQ1}) has been found by convex optimization. It can be seen that the results are in excellent numerical agreement with the LHS of (\ref{DoverQ1}), which were easily obtained in this paper and represented by cross marks.

\begin{figure}[t]
\begin{center}
  \includegraphics[width=\figwidth cm]{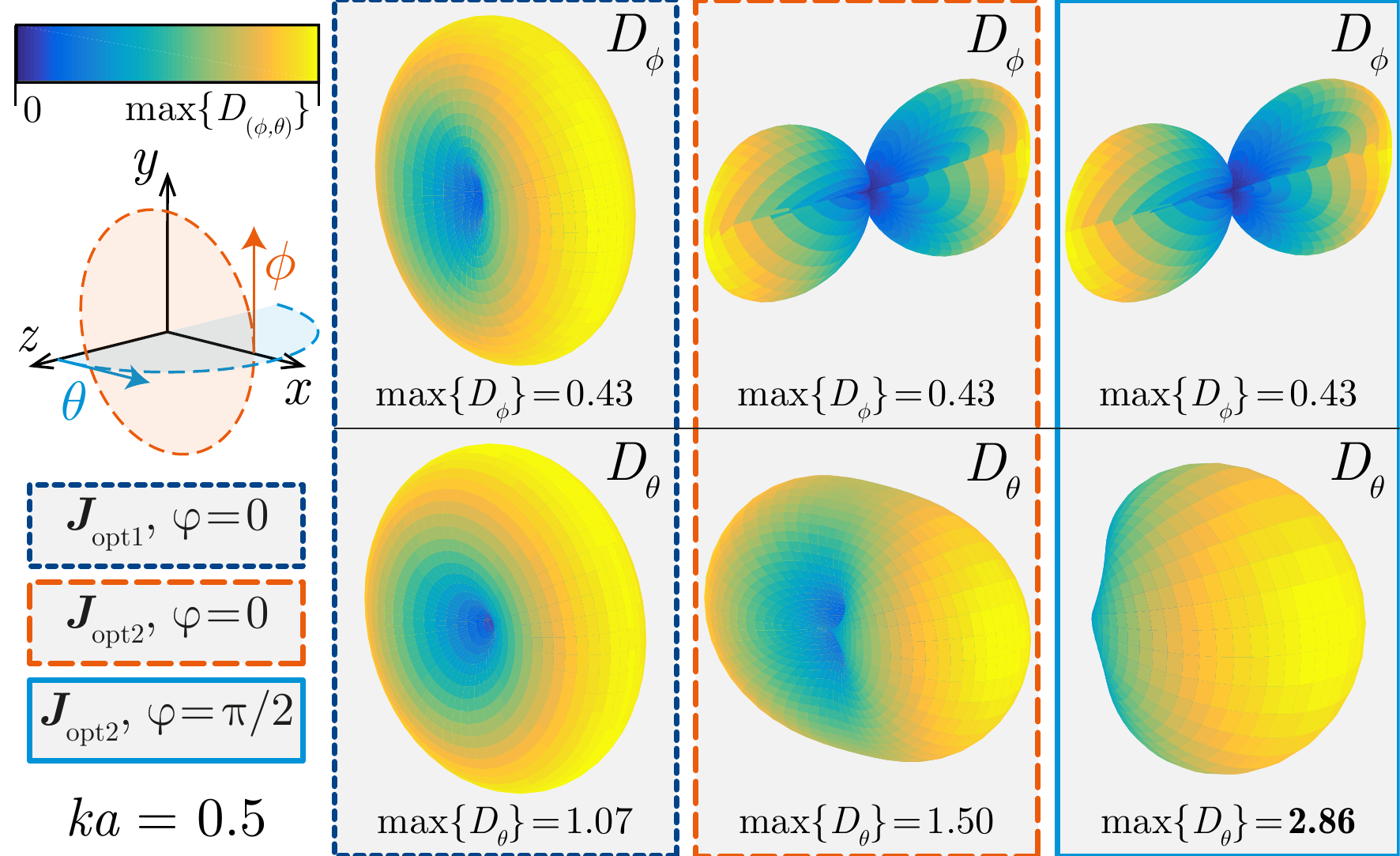} 
  \caption{The partial directivity $D_{\left(\phi,\theta\right)}$ for the currents $\M{J}_{\iopt 1}$ and $\M{J}_{\iopt 2}$ from Fig.~\ref{Fig5} which are optimal with respect to quality factor~$Q$. All results are depicted for a spherical shell of radius $a$ at $ka = 0.5$. Three boxes depict resulting directivity in terms of the dependence on the phase shift $\varphi$ between the dominant and the tuning mode. The left box shows the partial directivity for the current density $\M{J}_{\iopt 1}$ with both modes in phase. The resulting radiation pattern is omnidirectional \cite{Chu_PhysicalLimitationsOfOmniDirectAntennas}. The middle box and the right box show the partial directivity for the current density $\M{J}_{\iopt 2}$. In the middle, the modes are in phase and the maximal partial directivity is increased from $1.07$ to $1.50$. By setting the phase between modes to $\varphi = \pi / 2$, the directivity reaches a value of $2.86$, which is close to the theoretically predicted value of $3$ \cite{GustafssonCismasuJonsson_PhysicalBoundsAndOptimalCurrentsOnAntennas_TAP}.}
  \label{Fig10}
\end{center}
\end{figure}

In the case of the spherical shell, we have two solutions, \mbox{$\M{J}_{\iopt 1}$} and \mbox{$\M{J}_{\iopt 2}$}, minimizing quality factor~$Q$, as can be seen in Fig.~\ref{Fig6}. To find the value of the LHS in (\ref{DoverQ1}), the complete symmetry of a spherical shell is utilized so that the maximal partial directivity can freely be chosen
\BE
\label{DoverQ3}
\frac{\max\limits_{\theta,\phi}\left\{D_{\theta,\phi} \left(\Ivec_\iopt\right)\right\}}{\QT \left(\Ivec_\iopt\right)} \approx \max\limits_\Ivec \left\{ \frac{D_{\theta,\phi} \left(\Ivec\right)}{\QT \left(\Ivec\right)}\right\}
\EE
and the maximal partial directivity $D_{\theta,\phi}$ is evaluated for both \mbox{$\M{J}_{\iopt 1}$} and \mbox{$\M{J}_{\iopt 2}$} currents depicted in Fig.~\ref{Fig10}. In the context of the maximization the partial directivity, the phase shift $\varphi$ in $\alpha_\iopt$ (\ref{TCM10}) plays an important role. The highest directivity is obtained if the second current solution, \mbox{$\M{J}_{\iopt 2}$}, is taken and the modal currents are shifted by $\varphi = \pi /2$, as already anticipated in \cite{Chu_PhysicalLimitationsOfOmniDirectAntennas}, see Fig.~\ref{Fig10} for results at \mbox{$ka = 0.5$}. To recapitulate, all three solutions in Fig.~\ref{Fig10} minimize quality factor~$Q$, particularly \mbox{$\QT \left(\Ivec_\iopt\right) = 9.72$} for \mbox{$ka = 0.5$} gives \mbox{$(D_\theta / Q)/ka^3 = \left\{0.881, 1.234, 2.379\right\}$}. These results are consistent with the upper bound prediction $D/Q \leq 2.9 \left(ka\right)^3$ derived via the static polarizability approach \cite{GustafssonCismasuJonsson_PhysicalBoundsAndOptimalCurrentsOnAntennas_TAP}. However, as shown in Fig.~\ref{Fig7}, the predicted bound is reached only for small $ka$.

The perfect numerical agreement of the rectangle and the sphere can be extrapolated to the example of a circle. The circle represents the best uniplanar structure since it completely fills the $ka$ area in $x$-$y$ plane.

In the case of $G/Q$ optimization, the radiated power does not need to be maximized leaving us with a set of equally good solutions \cite{GustafssonCismasuJonsson_PhysicalBoundsAndOptimalCurrentsOnAntennas_TAP} tuned to the resonance by any non-radiating current \cite{GustafssonTayliEhrenborgCismasuNorbedo_MatlabCvxTutorial}. However, it is the additional requirement on minimal quality factor~$Q$ which reduces the set of possible solutions to one.

\section{Discussion}
\label{S4_Discussion}

Various properties and features of the method proposed above are discussed in this section.

\subsection{Optimality of the Surface}
\label{S41_SurfOpt}

Numerical results suggest that the simple connected surface filling the entire allocated space is always the best candidate to minimize quality factor~$Q$. If the outward boundary is preserved without any interruptions, quality factor $Q$ can be close to the minimal quality factor $Q$ of the original simple connected space, as shown in the comparison in Table~\ref{Table1_differentRectLineStrs}. A tiny difference between the rectangular plate and the loop of the same dimensions is reported. From this point, a trade-off between decreasing the electrical size and increasing quality factor~$Q$ of a self-resonant antenna can be understood from another perspective: to ensure the resonance, the surface needs to be appropriately deformed, however, the low quality factor~$Q$ dictates the space to be simple connected without any disturbances on which the charge can be accumulated \cite{YaghjianStuart_LowerBoundOnTheQofElectricallySmallDipoleAntennas}.

\begin{table}[t] 
\centering 
\begin{tabular}{|c||c|c|c|}
\hline 
$\Omega$ ($ka = 0.5$) & $\frac{\QT\left(\Ivec_1\right)}{\QchuTM}$ & $\frac{\QT\left(\Ivec_\iopt\right)}{\QT\left(\Ivec_1\right)}$ & $\frac{G_y}{\QT \left(\Ivec_\iopt\right)}$ \\
\hline 
\hline 
\noindent\parbox[c]{\tabwidth cm}{\includegraphics[width=\tabwidth cm, page=5]{Fig4_tuningPosibilities_Table.pdf}}
 & 4.250 & 0.839 & 0.0352 \\
\hline   
\noindent\parbox[c]{\tabwidth cm}{\includegraphics[width=\tabwidth cm, page=6]{Fig4_tuningPosibilities_Table.pdf}}
 & 4.301 & 0.840 & 0.0349 \\
\hline 
\noindent\parbox[c]{\tabwidth cm}{\includegraphics[width=\tabwidth cm, page=7]{Fig4_tuningPosibilities_Table.pdf}}
 & 4.344 & 0.842 & 0.0347 \\
\hline 
\noindent\parbox[c]{\tabwidth cm}{\includegraphics[width=\tabwidth cm, page=2]{Fig4_tuningPosibilities_Table.pdf}}
 & 4.399 & 0.839 & 0.0343 \\
\hline 
\noindent\parbox[c]{\tabwidth cm}{\includegraphics[width=\tabwidth cm, page=1]{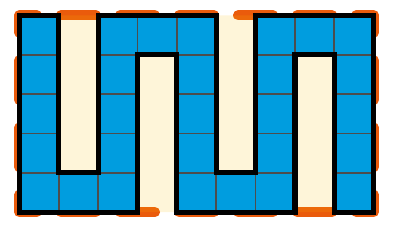}}
 & 4.420 & 0.995 & 0.0285 \\
\hline 
\noindent\parbox[c]{\tabwidth cm}{\includegraphics[width=\tabwidth cm, page=8]{Fig4_tuningPosibilities_Table.pdf}}
 & 4.670 & 1.008 & 0.0283 \\  
\hline 
\hline 
\end{tabular}
\caption{Rectangular plate and its five geometrical modifications of the electrical size $ka = 0.5$. The structures are depicted in the same scale and discretized with $\Nmeas = 14240 \pm 340$. The meaning and the evaluation of the gain $G_y$ in the fourth column is the same as in case of the dipole and the rectangle in Fig.~\ref{Fig7}.}
\label{Table1_differentRectLineStrs} 
\end{table}

\subsection{Selection of the Modes}
\label{S42_ModeSel}

Generally, the tuning made by only two modal currents does not automatically ensure minimal quality factor $Q$. However, once the cross-terms \mbox{$\Ivec_u^\herm \XD \Ivec_v$} can be neglected ($ka < 1$) or they are not present, as in (\ref{XM1}), the optimum can always be approached by only two properly chosen modes. 

It was already demonstrated that quality factor of the dominant mode~$\QT\left(\Ivec_1\right)$ cannot always be reduced by adding the second mode, see the fractal shape in Fig.~\ref{Fig1}. Notice, however that we can always use/imagine a mode with an extremely high eigenvalue \cite{GustafssonTayliEhrenborgCismasuNorbedo_MatlabCvxTutorial} that serves similarly as a tuning element (it has practically zero radiated power and high magnetic or electric energy). This current will probably be extremely spatially localized, thus, difficult to find in the characteristic mode basis. 

\begin{figure}[t]
\begin{center}
  \includegraphics[width=\figwidth cm]{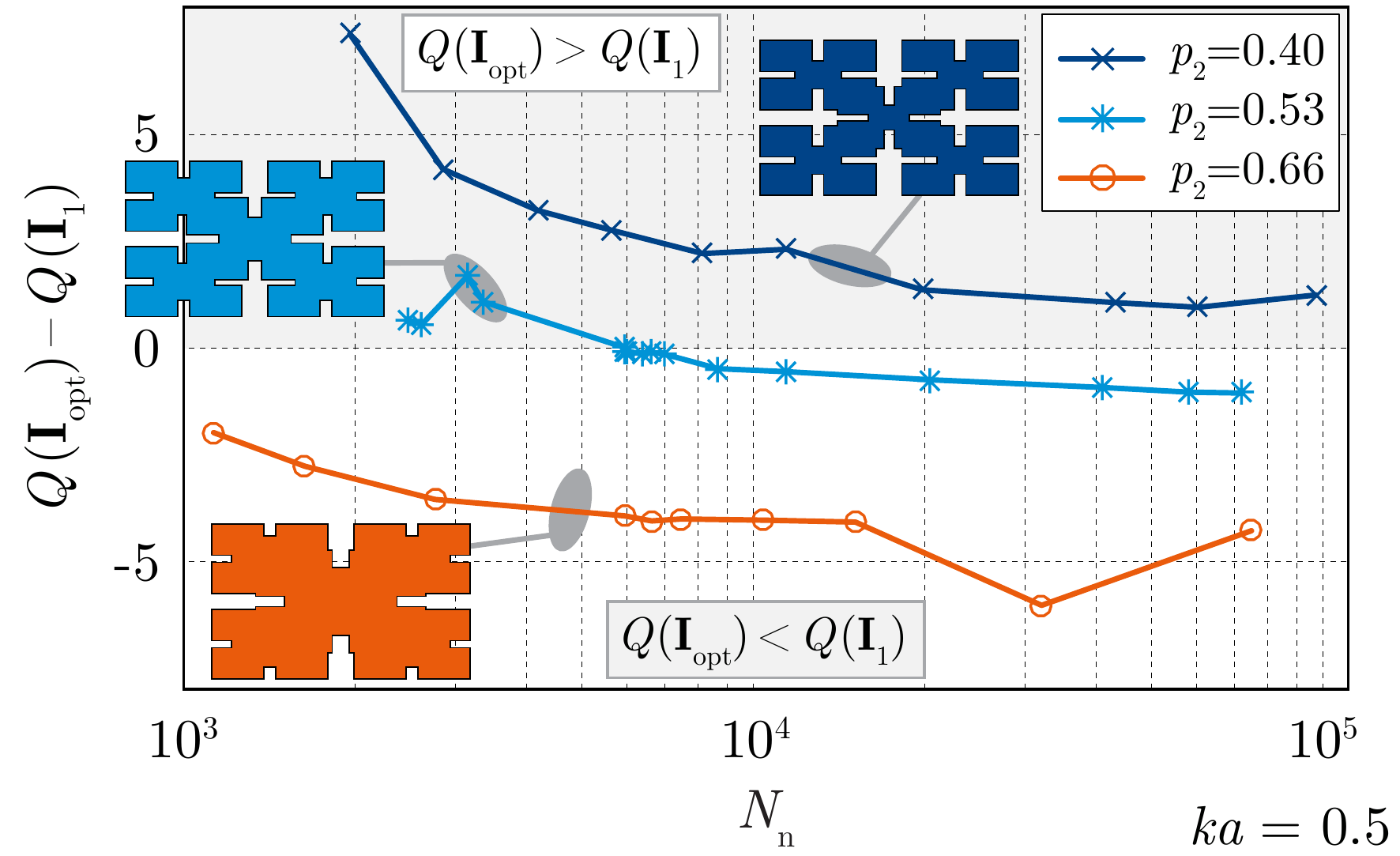}
  \caption{Three slightly different perfectly conducting IFS fractal structures \cite{Falconer_FractalGeometry} of second iteration with electrical size $ka = 0.5$. The geometrical complexity is controlled by parameter $p_2$ whose meaning is explained in Appendix~\ref{appB}. All three structures have similar quality factor \mbox{$\QT \left(\Ivec_1\right) = 42\pm3$} but completely dissimilar tuning potential given by the tuning (inductive) mode. The most complex shape (dark blue) cannot be tuned for any $\Nmeas$. On the other hand, the simplest shape (orange) can always be tuned. A trade-off between these extrema is shape (light blue) tunable in dependence on $\Nmeas$. Regardless of the tuning potential, all structures can always be tuned by an external lumped element or, alternatively, by a localized current.}
  \label{Fig6}
\end{center}
\end{figure}

To this point, it is useful to check the condition (\ref{TCMopt1}) if the mode doublet $\Ivec_1$ and $\Ivec_v$ is capable of reducing the quality factor of the dominant mode $\QT\left(\Ivec_1\right)$. The test is performed on three similar structures of different geometrical complexity in Fig.~\ref{Fig6}. The role of the discretization is crucial as the tuning possibilities (\ref{TCMopt1}) depend not only on $\Omega$ and $ka$, but also on $\Nmeas$. It is necessary to sufficiently mesh the structure, e.g., the tuning mode (often first inductive mode) needs to be represented correctly. Practically, it means that at least 2-3 segments must be used in the narrowest part of the structure.

Collecting the observations from the two recent sections, the generally applicable work-flow (PEC, $ka < 1$, $\int_\Omega \D{V} = 0$) is as follows:
\begin{itemize}
\item determine the maximal available surface region,
\item make it simple connected,
\item solve (\ref{TCM1}) or (\ref{XM1}),
\item determine the proper capacitive-inductive mode doublet with the lowest $\QU$,
\item use (\ref{TCM10}) or (\ref{XM6}).
\end{itemize}

\subsection{Numerical Issues}
\label{S44_NumIssues}

The procedure deals only with the first several modes which have the lowest eigenvalue. For this reason the Iteratively Restarted Arnoldi (IA) method can safely be used, searching only for a few modes. The IA algorithm has propitious algorithmic complexity \mbox{$\propto\mathcal{O} \left(m U^2\right)$} where $m$ is the number of desired modes (2-5 in this paper). Notice the speed of the IA method in comparison to the generalized Schur (gS) decomposition \cite{Wilkinson_AlgebraicEigenvalueProblem} of complexity \mbox{$\propto \mathcal{O} \left(U^3\right)$}. Also, the numerical issues, associated with the higher modes, are of no concern here.

\subsection{Volumetric Antennas}
\label{S45_VolAntennas}

A technique showing how to reduce the quality factor~$Q$ of volumetric radiator by combining the electric and magnetic currents, which has as been recently published \cite{Kim_LowerBoundsOnQForFinizeSizeAntennasOfArbitraryShape}, can advantageously be combined with the approach presented in this paper. Now, as a first guess, the best electric current can be used instead of the dominant characteristic mode \cite{Kim_LowerBoundsOnQForFinizeSizeAntennasOfArbitraryShape}. Since the electric currents are combined in this work, it perfectly suits all surface shapes. On the other hand, for volumetric structures like a spherical shell, the method from \cite{Kim_LowerBoundsOnQForFinizeSizeAntennasOfArbitraryShape} yields better results since it is able to exclude the stored energy from the interior of the antenna.

\subsection{Open Questions}
\label{S46_Questions}

The proposed method also poses several interesting, yet unsolved, questions:
\begin{enumerate}
\item How are the optimal currents related to the current which can be excited by realistic feeding \cite{MartensManteuffel_SystematicDesignMethodOfMobileMultipleAntennaSystemUsingCM}? One possible way to synthesise the feeding network is to try to eliminate all higher modes (e.g., by adding slots to the proper places without causing a significant impact on the quality factor~$Q$ or on the $G/Q$ ratio) and to feed the dominant modes at the appropriate places \cite{CapekHazdraEichler_AMethodForTheEvaluationOfRadiationQBasedOnModalApproach}. Notice, in this context, that for characteristic modes \cite{HarringtonMautz_TheoryOfCharacteristicModesForConductingBodies}, the summation formula (\ref{ModQ3}) is directly related to the external world represented by feeding
\BE
\label{Disc1}
\M{I} = \sum\limits_{u = 1}^U \frac{\Ivec_u^\trans \Vi}{1 + \J \lambda_u} \frac{\Ivec_u}{\Ivec_u^\herm \M{R}\Ivec},
\EE
where $\Vi$ are the excitation coefficients \cite{HarringtonMautz_TheoryOfCharacteristicModesForConductingBodies}.

\item The question closely related to the previous one is a trade-off between the smallest area used and the resulting quality factor~$Q$ or the $G/Q$ ratio. This problem can hardly be solved by a non-heuristic optimization approach since it involves geometry modifications. However, the proposed technique offer useful procedure how to do that -- it is sufficient to study only quality factor~$\QU$ of the dominant mode and the tuning potential of the tuning mode.

\item The ohmic losses are recognized as one of the determining phenomena for the ESA. The solution in this paper is found through well-radiating modes and, thus, the ohmic losses can, potentially, be minimized as they are equal to the magnitude of the current \cite{CapekEichlerHazdra_EvaluationOfRadiationEfficienciIET}. These questions on minimizing the ohmic losses remain unanswered.
\end{enumerate}

\section{Conclusion}
\label{Concl}

The complex optimization problem of isolating the resonant current with the minimal quality factor $Q$ has been investigated. A simple procedure to solve the problem for electrically small structures has been proposed, utilizing the modal decomposition of the underlying matrix operators.

Since the stored energy cross-terms between the modal currents have been neglected, the solutions are approximate with respect to the fundamental bounds on quality factor~$Q$ and the \mbox{$G/Q$} ratio. However, the procedure is near exact for electrically small antennas as the cross-terms are truly negligible.

It has been observed that the dominant mode can always be tuned by another modal current instead of the conventional tuning procedure made by an external lumped element. The exact condition under which quality factor~$Q$ can be significantly lowered has been derived. The presented approach is straightforward, needs only one solution of the generalized eigenvalue problem, and, once the proper current basis is known, the problem can be, surprisingly, solved analytically. The procedure is not dependent on a particular choice of the space discretization and the used basis functions.

The technique has been verified on a number of examples, including a rectangular plate, a spherical shell and other planar structures of varying geometrical and topological complexity. A number of already known results have been confirmed, which verifies the development undertaken. Beyond this, new results have been presented which can be utilized by other researchers. To give but one example, the resulting electric current can be employed as a first guess if the quality factor~$Q$ is minimized by a combination of electric and magnetic currents.

New questions and new directions as to analyse electrically small antennas are now available since a log road must be undertaken to bridge the gap between optimal current isolation and the capability to fabricate optimal antennas. The various shape modifications can be effectively studied and compared with the original simple connected regions with lower quality factor~$Q$. This task involves adding slots and the feeders so that a current similar to the optimal one will be excited. Another avenue of investigation can be directed at studying ohmic losses and, in particular, modal losses.

\appendices
%

\section{Proof of Minimal Quality Factor $Q$ for $\Qext = 0$}
\label{appCM1}
Proof that the minimal quality factor $\QT \left(\Ivec_\iopt\right) = \Qmin$ occurs at self-resonance of the current $\Ivec_\iopt$ is given in this appendix.

Generally, two options, denoted as $\QT^\mathrm{A}$ and $\QT^\mathrm{B}$, can appear: $\Qmin$ occurs at self-resonance of the optimal current, $\Qext = 0$, or $\Qmin$ occurs outside self-resonance, $\Qext \neq 0$. Hereafter, assume that no cross-terms are presented.
\begin{itemize}
\item If $\Qext = 0$, then
\BE
\label{QminProof1}
\QT^\mathrm{A} \left(\Ivec_\iopt\right) = \frac{\Ivec_\iopt^\herm \XM \Ivec_\iopt + \Ivec_\iopt^\herm \XE \Ivec_\iopt}{4 \left( 1 + \left|\alpha_\iopt\right|^2\right)}.
\EE
\item If $\Qext \neq 0$, then, without loss of generality, let us suppose that the electric part of the stored energy predominates the magnetic part,
\BE
\label{QminProof2}
\Ivec_\iopt^\herm \XE \Ivec_\iopt > \Ivec_\iopt^\herm \XM \Ivec_\iopt.
\EE
Then
\BE
\label{QminProof3}
\QT^\mathrm{B} \left(\Ivec_\iopt\right) = \frac{\Ivec_\iopt^\herm \XE \Ivec_\iopt}{2 \left( 1 + \left|\alpha_\iopt\right|^2\right)}.
\EE
\end{itemize}
Now, if we subtract the B possibility (\ref{QminProof3}) from the A possibility (\ref{QminProof1}), we get
\BE
\label{QminProof4}
\QT^\mathrm{B} - \QT^\mathrm{A} = \frac{\Ivec_\iopt^\herm \XE \Ivec_\iopt - \Ivec_\iopt^\herm \XM \Ivec_\iopt}{4 \left( 1 + \left|\alpha_\iopt\right|^2\right)},
\EE
and since (\ref{QminProof2}) has been assumed, both sides of (\ref{QminProof4}) are always positive, which indicates that 
\BE
\label{QminProof5}
\QT^\mathrm{B} > \QT^\mathrm{A}.
\EE
The expression (\ref{QminProof5}) leads to the clear conclusion that the minimal quality factor~$Q$ is always obtained for a current $\Ivec_\iopt$ which is self-resonant.

\section{Reduction of the Modal Quality Factor~$\QT\left(\Ivec_1\right)$}
\label{appCM2}

The purpose of this appendix is to derive the condition
\BE
\label{appCM1ex1}
\QT\left(\Ivec_\iopt\right) < \QT\left(\Ivec_1\right)
\EE
in terms of the eigenvalues and the modal stored energies. The formula (\ref{appCM1ex1}) is expressed according to (\ref{ModQ4})
\BE
\label{appCM1ex2}
\frac{\Ivec_1^\herm \XD \Ivec_1 + \left|\alpha_\iopt\right|^2 \Ivec_2^\herm \XD \Ivec_2}{2 \left(1 + \left|\alpha_\iopt\right|^2\right)} < \frac{\Ivec_1^\herm \XD \Ivec_1 + \left|\Ivec_1^\herm \M{X} \Ivec_1\right|}{2},
\EE
then expression (\ref{TCM9}) is used
\BE
\label{appCM1ex3}
\Ivec_2^\herm \XD \Ivec_2 - \Ivec_1^\herm \XD \Ivec_1 < \frac{\left|\lambda_1\right| + \left|\lambda_2\right|}{\left|\lambda_1\right|} \left|\Ivec_1^\herm \M{X} \Ivec_1\right|.
\EE
As a last step, (\ref{TCM3}) is substituted into (\ref{appCM1ex3}) leaving us with
\BE
\label{appCM1ex4}
\Ivec_2^\herm \XD \Ivec_2 - \Ivec_1^\herm \XD \Ivec_1 < \left|\lambda_1\right| + \left|\lambda_2\right|.
\EE
Remember, the modal radiated power has been normalized according to (\ref{TCM2}).

\section{Characteristic Numbers Related to Spherical Harmonics}
\label{appA}
The characteristic numbers $\lambda_\mathrm{TM10}$ and $\lambda_\mathrm{TE10}$ of the dominant $\text{TM}_{10}$ and $\text{TE}_{10}$ characteristic modes on a spherical shell are found in this appendix by an analogy with spherical harmonics. 

Using (\ref{StoredEnergy3A}), (\ref{StoredEnergy3B}) and (\ref{QualityFactor2}), the (\ref{TCM3}) can be rewritten as
\BE
\label{CML1}
\lambda = \frac{2 \omega\left(\Wm - \We\right)}{\Prad}.
\EE
The quantities $\Prad$, $\Wm$ and $\We$ in (\ref{CML1}) can be found for any separable system analytically, see \cite{CapekJelinekHazdraEichler_QofSphere} for a particular case of a sphere. Substituting (11a)--(11d) and (17a)--(17b) from \cite{CapekJelinekHazdraEichler_QofSphere} into (\ref{CML1}), we immediately get
\begin{subequations}
\begin{align}
\label{CML3A}
\lambda_\mathrm{TM10} &= - \frac{\YBSL{1}{ka} - ka \, \YBSL{0}{ka}}{\JBSL{1}{ka} - ka \, \JBSL{0}{ka}}, \\
\label{CML3B}
\lambda_\mathrm{TE10} &= - \frac{\YBSL{1}{ka}}{\JBSL{1}{ka}},
\end{align}
\end{subequations}
where $\JBSL{n}{ka}$ and $\YBSL{n}{ka}$ are spherical Bessel functions of the $n$th order and of the first and the second kind, respectively. The derivation for the higher-order modes is left to the reader.

\section{Iterated Function System -- Fractal Generation}
\label{appB}

The process of generating IFS (iterative function system) fractals is described in this appendix.

IFS is a finite set of contraction mappings on a complete metric space \cite{Falconer_FractalGeometry}. It can be generated with the Hutchinson operator \cite{Edgar_MeasureTopologyAndFractalGeometry}:
\BE
\label{IFS1}
\Omega_{n+1} = \bigcup\limits_i f_i \left(\Omega_n\right),
\EE
in which $f_i$ is the $i$th affine transformation and $n$ denotes the iteration. For particular shapes given in this paper, the initial shape  is a rectangle centred in the middle of the coordinate system and aligned with $x$-$y$ axes. Its length is $L$ and width is $3 L / 5$. The affine transformations $f_i$ are applied to each point $\M{P}$ of $\Omega_n$. They are defined by six coefficients as follows:
\BE
\label{IFS2}
f_i: \M{P}_{n+1} = \left( \begin{array}{cc}
c_1 & c_2 \\
c_3 & c_4 \\
\end{array} \right)
\M{P}_n + 
\left( \begin{array}{c}
c_5 \\
c_6 \\
\end{array} \right),
\EE
the particular values can be found in Table~\ref{Table2_IFS}. Two iterations, $n=2$, of (\ref{IFS1}) have been applied.
\begin{table}[t] 
\centering 
\begin{tabular}{|c|c|c|c|c|c|c|}
\hline  & $c_1$ & $c_2$ & $c_3$ & $c_4$ & $c_5$ & $c_6$ \\ 
\hline
\hline $f_1$ & $p_1$ & 0 & 0 & $p_1$ & $\frac{1}{2} \left(1 - p_1\right) L$ & $\frac{3}{10} \left(1 - p_1 \right) L$ \\ 
\hline $f_2$ & $p_1$ & 0 & 0 & $p_1$ & $-\frac{1}{2} \left(1 - p_1\right) L$ & $\frac{3}{10} \left(1 - p_1 \right) L$ \\ 
\hline $f_3$ & $p_1$ & 0 & 0 & $p_1$ & $-\frac{1}{2} \left(1 - p_1\right) L$ & $-\frac{3}{10} \left(1 - p_1 \right) L$ \\ 
\hline $f_4$ & $p_1$ & 0 & 0 & $p_1$ & $\frac{1}{2} \left(1 - p_1\right) L$ & $-\frac{3}{10} \left(1 - p_1 \right) L$ \\ 
\hline $f_5$ & $p_2$ & 0 & 0 & $p_2$ & 0 & 0  \\ 
\hline 
\hline 
\end{tabular} 
\caption{The affine transformations for the generation of fractal shapes used in the paper. The contractions are $p_1 = 0.45$, $p_2 = \left\{ 0.2, 0.4, 0.53, 0.66 \right\}$.} 
\label{Table2_IFS} 
\end{table} 

\section{Directivity matrix}
\label{appC}

The procedure to find the far-field vectors $\M{F}_\theta$ and $\M{F}_\phi$ is presented in \cite{GustafssonTayliEhrenborgCismasuNorbedo_MatlabCvxTutorial}. To get the directivity as a bilinear form, the matrix representation of the directivity operator is found as
\begin{subequations}
\begin{align}
\label{Dmatrix1A}
\M{D}_\theta &= \frac{2\pi}{\Zvac}\frac{\M{F}_\theta^\herm \M{F}_\theta}{\Prad} = \frac{4\pi}{\Zvac}\frac{\M{F}_\theta^\herm \M{F}_\theta}{\Ivec^\herm \M{R} \Ivec}, \\
\label{Dmatrix1B}
\M{D}_\phi &= \frac{2\pi}{\Zvac}\frac{\M{F}_\phi^\herm \M{F}_\phi}{\Prad} = \frac{4\pi}{\Zvac}\frac{\M{F}_\phi^\herm \M{F}_\phi}{\Ivec^\herm \M{R} \Ivec},
\end{align}
\end{subequations}
for $\theta$ and $\phi$ components, respectively, and the total directivity matrix reads
\BE
\label{Dmatrix2}
\M{D} = \M{D}_\theta + \M{D}_\phi.
\EE
The directivity is then calculated as
\BE
\label{Dmatrix3}
D_{\left(\theta / \phi\right)} =  \Ivec^\herm \M{D}_{\left(\theta / \phi\right)} \Ivec.
\EE

\section*{Acknowledgement}
The authors would like to thank Mats Gustafsson from the University of Lund (Sweden) for his thoughts and suggestions which stimulated the development of the presented work.

\ifCLASSOPTIONcaptionsoff
  \newpage
\fi

\bibliographystyle{IEEEtran}
\bibliography{references_LIST}

\begin{biography}[{\includegraphics[width=1in,height=1.25in,clip,keepaspectratio]{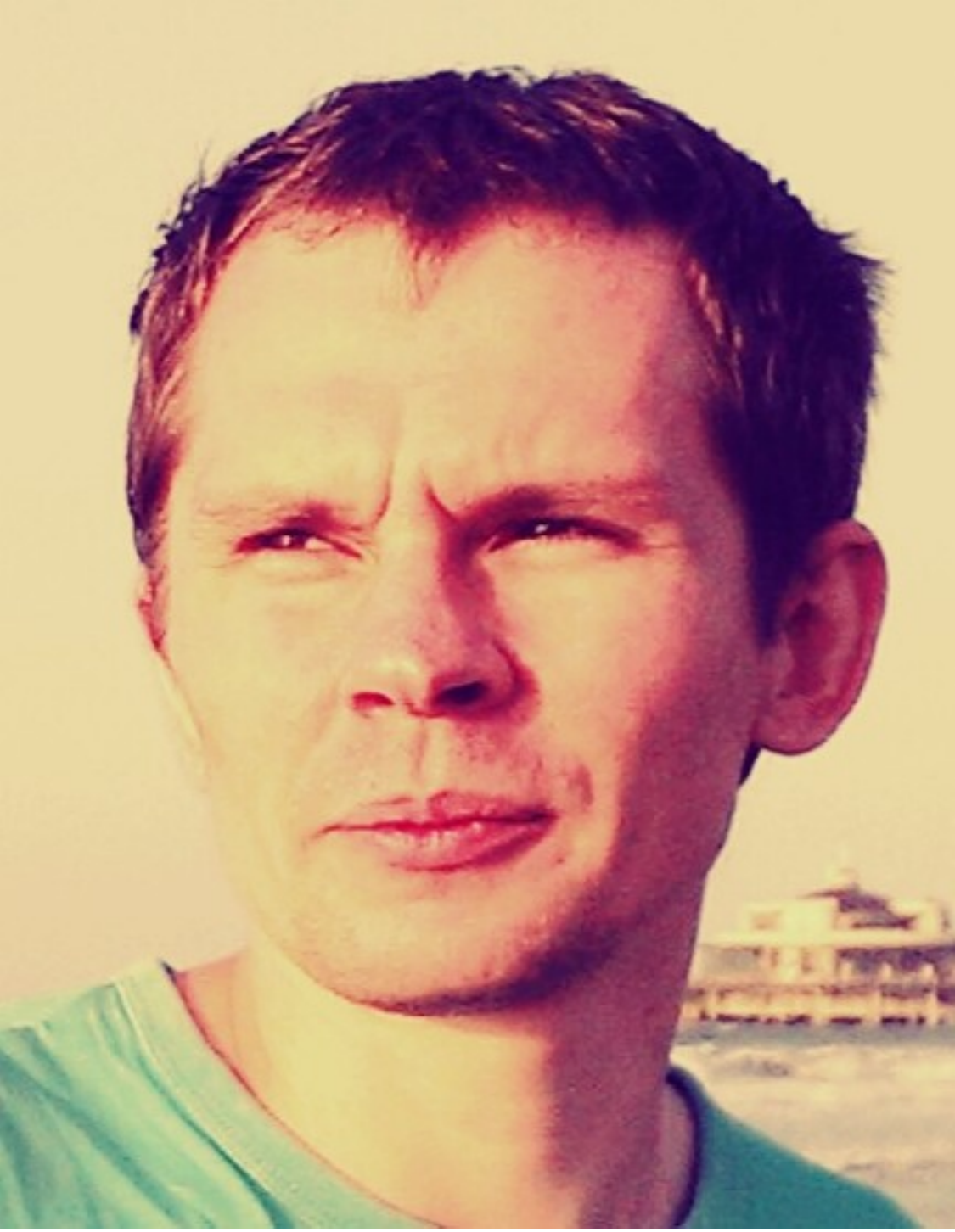}}]{Miloslav Capek}
(S'09, M'14) received his M.Sc. degree in Electrical Engineering from the Czech Technical University, Czech Republic, in 2009, and his Ph.D. degree from the same University, in 2014. Currently, he is a researcher with the Department of Electromagnetic Field, CTU-FEE.
	
He leads the development of the AToM (Antenna Toolbox for Matlab) package. His research interests are in the area of electromagnetic theory, electrically small antennas, numerical techniques, fractal geometry and optimization.

\end{biography}
\begin{biography}[{\includegraphics[width=1in,height=1.25in,clip,keepaspectratio]{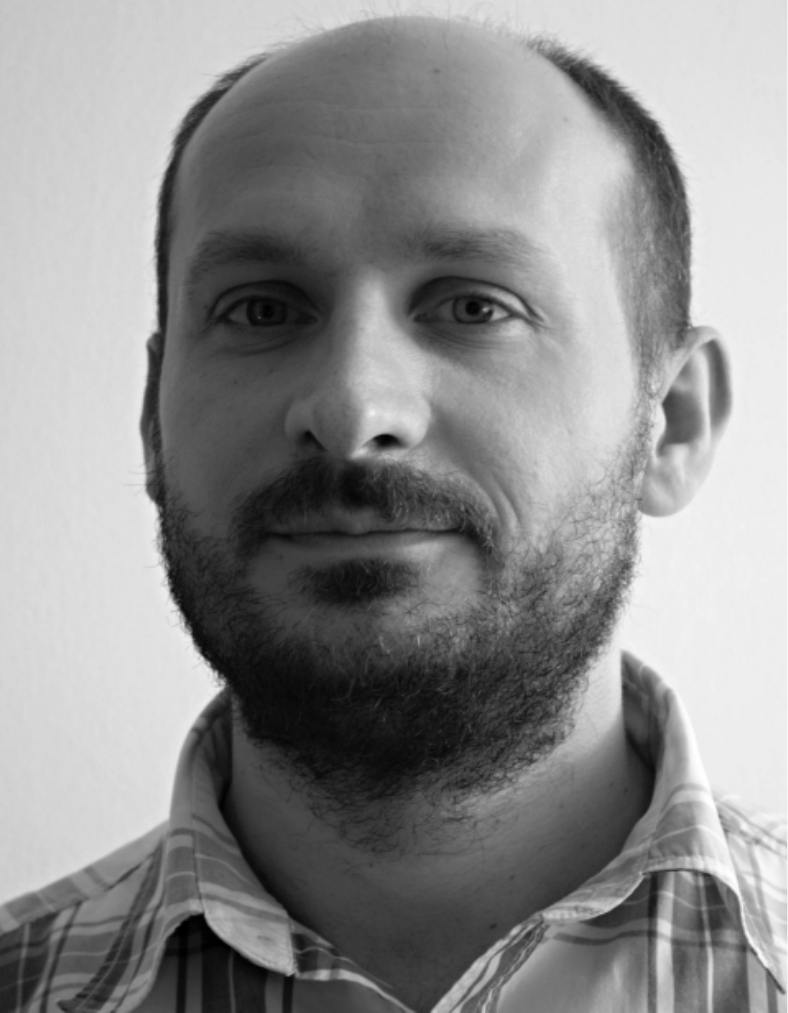}}]{Lukas Jelinek}
received his Ph.D. degree from the Czech Technical University in Prague, Czech Republic, in 2006. In 2015 he was appointed Associate Professor at the Department of Electromagnetic Field at the same university.

His research interests include wave propagation in complex media, general field theory, numerical techniques and optimization.
\end{biography}
\end{document}